\documentclass[lettersize,journal]{IEEEtran} 
\usepackage{amsmath,amsfonts}
\usepackage{algorithmic}
\usepackage{algorithm}
\usepackage{array}
\usepackage[caption=false,font=normalsize,labelfont=sf,textfont=sf]{subfig}
\usepackage{textcomp}
\usepackage{stfloats}
\usepackage{url}
\usepackage{verbatim}
\usepackage{graphicx}
\usepackage{cite}

\usepackage{soul}
\usepackage{pgfplotstable}
\usepackage{pgfplots, tikz}
\pgfplotsset{compat=1.17}
\usepackage{microtype, booktabs}

\usepackage{calrsfs}
\DeclareMathAlphabet{\pazocal}{OMS}{zplm}{m}{n}
\newcommand{\La}{\pazocal{L}}

\usepackage{hyperref}

\usepackage{tablefootnote}

\begin{document}

\title{A Neural-Network Framework for the Design of Individualised Hearing-Loss Compensation}

\author{Fotios Drakopoulos,~\IEEEmembership{Member,~IEEE,} Sarah Verhulst,~\IEEEmembership{Member,~IEEE} 
        
\vspace{-20pt}

\thanks{The authors are with the Hearing Technology Lab, Department of Information Technology, Ghent University, 9000 Ghent, Belgium (e-mail: f.drakopoulos@ucl.ac.uk; s.verhulst@ugent.be). Corresponding author: Fotios Drakopoulos (present address: UCL Ear Institute, University College London, WC1E 6BT London, U.K.).

The authors declare the following competing interests: A patent application (WO2021198438) was filed by Ghent University (UGent) on the basis of the research presented in this manuscript. This work was supported by the European Research Council (ERC) under the Horizon 2020 Research and Innovation Programme (grant agreement No 678120 RobSpear). 

The code for using the CoNNear auditory periphery model is available via \protect\url{https://doi.org/10.5281/zenodo.6834691} or \protect\url{https://github.com/HearingTechnology/CoNNear_periphery}. A non-commercial, academic UGent license applies.}} 



\maketitle

\begin{abstract}
Sound processing in the human auditory system is complex and highly non-linear, whereas hearing aids (HAs) still rely on simplified descriptions of auditory processing or hearing loss to restore hearing. Even though standard HA amplification strategies succeed in restoring audibility of faint sounds, they still fall short of providing targeted treatments for complex sensorineural deficits and adverse listening conditions.
These shortcomings of current HA devices demonstrate the need for advanced hearing-loss compensation strategies that can effectively leverage the non-linear character of the auditory system. 
Here, we propose a differentiable deep-neural-network (DNN) framework that can be used to train DNN-based HA models based on biophysical auditory-processing differences between normal-hearing and hearing-impaired systems.
We investigate different loss functions to accurately compensate for impairments that include outer-hair-cell (OHC) loss and cochlear synaptopathy (CS), and evaluate the benefits of our trained DNN-based HA models for speech processing in quiet and in noise.
Our results show that auditory-processing enhancement was possible for all considered hearing-loss cases, with OHC loss proving easier to compensate than CS.
Several objective metrics were considered to estimate the expected speech intelligibility after processing, and these simulations hold promise in yielding improved understanding of speech-in-noise for hearing-impaired listeners who use our DNN-based HA processing.
Since our framework can be tuned to the hearing-loss profiles of individual listeners, we enter an era where truly individualised and DNN-based hearing-restoration strategies can be developed and be tested experimentally.
\end{abstract}

\begin{IEEEkeywords}
neural networks, closed loop, individualised hearing aids, hearing loss, cochlear synaptopathy, speech intelligibility
\end{IEEEkeywords}


\vspace{-10pt}
\section{Introduction}
\IEEEPARstart{C}{omputational} models of hearing-impaired (HI) auditory processing are widely adopted in the design of hearing-restoration strategies \cite{byrne1986national,scollie2005desired,keidser2011nal,kates2014hearing,moore2010development}, but the numerical complexity of biophysically realistic auditory models has discouraged the field to make use of their full potential \cite{vecchi2021comparative}. Consequently, typical hearing-aid (HA) algorithms still rely on simplified descriptions of auditory processing or on superficial estimates of sensorineural hearing loss (SNHL) to yield optimal acoustic amplification to HA users \cite{smeds2015proprietary,marriage2018comparison}.
Although standard amplification strategies succeed in restoring the audibility of faint sounds in HA users based on measured hearing thresholds or perceived loudness, they fall short of providing a robust and generalisable treatment, as HA treatment outcomes are highly variable across listeners even with similar audiograms \cite{lopez2017predictors}.
These mixed benefits of modern hearing healthcare have led to a growing demand for fundamental advances in hearing research and more precise SNHL treatment outcomes \cite{lesica2021harnessing}.
To this end, more complex computational tools that can leverage the highly non-linear character of the human auditory system (and its impairments) are necessary to improve the diagnosis and treatment of hearing loss.

Advanced computational techniques based on deep neural networks (DNNs) have the potential to transform hearing healthcare and have already been incorporated in HA devices, substantially improving the understanding of speech in noise for HA users \cite{slaney2020auditory, healy2020talker}.
However, the specific functions of the auditory system have only partially been considered in the design of DNN-based HA algorithms and DNNs are mostly adopted to improve noise suppression, source separation or amplification fittings in HA technologies \cite{slaney2020auditory, healy2020talker, tu2021dhasp, nagathil2021computationally}, rather than focussing on the hearing-loss compensation per se.
To bridge this gap between biophysical plausibility and HA processing, computational closed-loop systems based on biophysically realistic auditory models have been proposed \cite{bondy2004novel, chen2005novel, haykin2006binaural}, with the aim of ``reverse-engineering'' HI auditory processing to optimise HA sound processing. 
However, the highly non-linear and non-differentiable nature of such descriptions makes it impossible to derive robust and generalisable solutions using gradient-free optimisation methods.
In contrast, DNN-based auditory models can overcome these limitations by providing differentiable descriptions of human auditory processing, such that the gradient of a closed-loop system can be computed analytically and be used by gradient-based optimisation algorithms (e.g. backpropagation) to achieve precise hearing-loss compensation.

To address this challenge and advance current HA technologies, we present a differentiable closed-loop framework for hearing-loss compensation that is based on a convolutional-neural-network (CNN) description of (impaired) human auditory processing \cite{baby2021convolutional,Drakopoulos2021}.
Unlike traditional HA signal processing, the differentiable nature of the adopted CNN-based model (CoNNear) allows DNN-based HA (DNN-HA) models to be trained within our framework from the ground up, without posing any prior constraints on the applied signal processing (e.g. frequency analysis/synthesis, pre-defined gain/compression functions).
At the same time, CoNNear was derived from a biophysically realistic analytical model of normal and impaired human auditory processing \cite{verhulst_hearres2018} and comprises accurate descriptions of the cochlea, inner-hair-cell (IHC) and auditory-nerve-fibre (ANF) processing stages.
Thus, our framework includes a detailed description of the functional elements that constitute the human peripheral auditory system to target specific aspects of SNHL, with the aim of restoring simulated auditory-nerve (AN) responses of a HI system to the reference response of a normal-hearing (NH) system. 
Assuming that a perfect reconstruction of a simulated HI AN response to the NH response will result in restored hearing,
our preliminary studies \cite{Drakopoulos2021,drakopoulos2022differentiable} demonstrated how auditory-processing differences between NH and HI models can be minimised via backpropagation.

The backbone of our framework was first presented in \cite{drakopoulos2022differentiable} and adopts separate CoNNear modules to simulate each auditory-processing stage (Section~\ref{sec:hi}). Thus, specific hearing deficits can be introduced in the CoNNear HI periphery to train individualised DNN-HA models that compensate for different combinations of OHC loss as well as damage to the AN synapses, i.e. cochlear synaptopathy (CS) \cite{kujawa2009adding, sergeyenko2013age, furman2013noise, parthasarathy2019age}. 
To our knowledge, this is the first differentiable closed-loop framework that is based on biophysically realistic auditory models and can compensate for CS.
In this work, we present our framework in full detail and introduce a faster implementation of the CoNNear auditory model (Section~\ref{sec:branched}). We also take a next step by investigating which loss functions benefit speech processing for HI listeners, with a focus on restoring biophysically relevant auditory-processing aspects that are impacted by SNHL. Finally, we examine which objective metrics can be used to predict the benefits that our trained DNN-HA models are expected to have on perceptual aspects such as perceived quality or recognition of speech in noise.
Even though a variety of objective metrics exist \cite{van2018evaluation, schadler2018objective}, most of the proposed speech-intelligibility metrics fail to generalise to new data or non-linear processing strategies \cite{jorgensen2011predicting, yamamoto2020gedi} and cannot predict measured speech recognition well across diverse signal degradations \cite{van2018evaluation, li2021multi}.
At the same time, few studies have focussed on the intelligibility benefit prediction for HA processing \cite{falk2015objective, kates2018using, schadler2018objective, graetzer2021clarity}, making it difficult to hypothesise which of the existing objective metrics will be most suitable for the evaluation of our non-linear DNN-HA strategies.

In the next sections, we describe our DNN-based framework and adopt it to design an optimal HA model that compensates for a mixed SNHL pathology, consisting of a sloping high-frequency audiogram and a $>$50\% loss of ANFs. We introduce and compare several loss functions that focus on the optimisation of different aspects of HI auditory processing, to identify the DNN-HA model that yields the best enhancement of simulated AN responses, while offering improved objective perceptual scores. 
To investigate the extent to which our methodology is applicable to other types of SNHL, we also apply the same training procedure for hearing-loss profiles that only comprise OHC loss or CS.

\section{Framework}

\begin{figure*}[tb!]
\begin{center}
\includegraphics[width=\textwidth,height=\textheight,keepaspectratio]{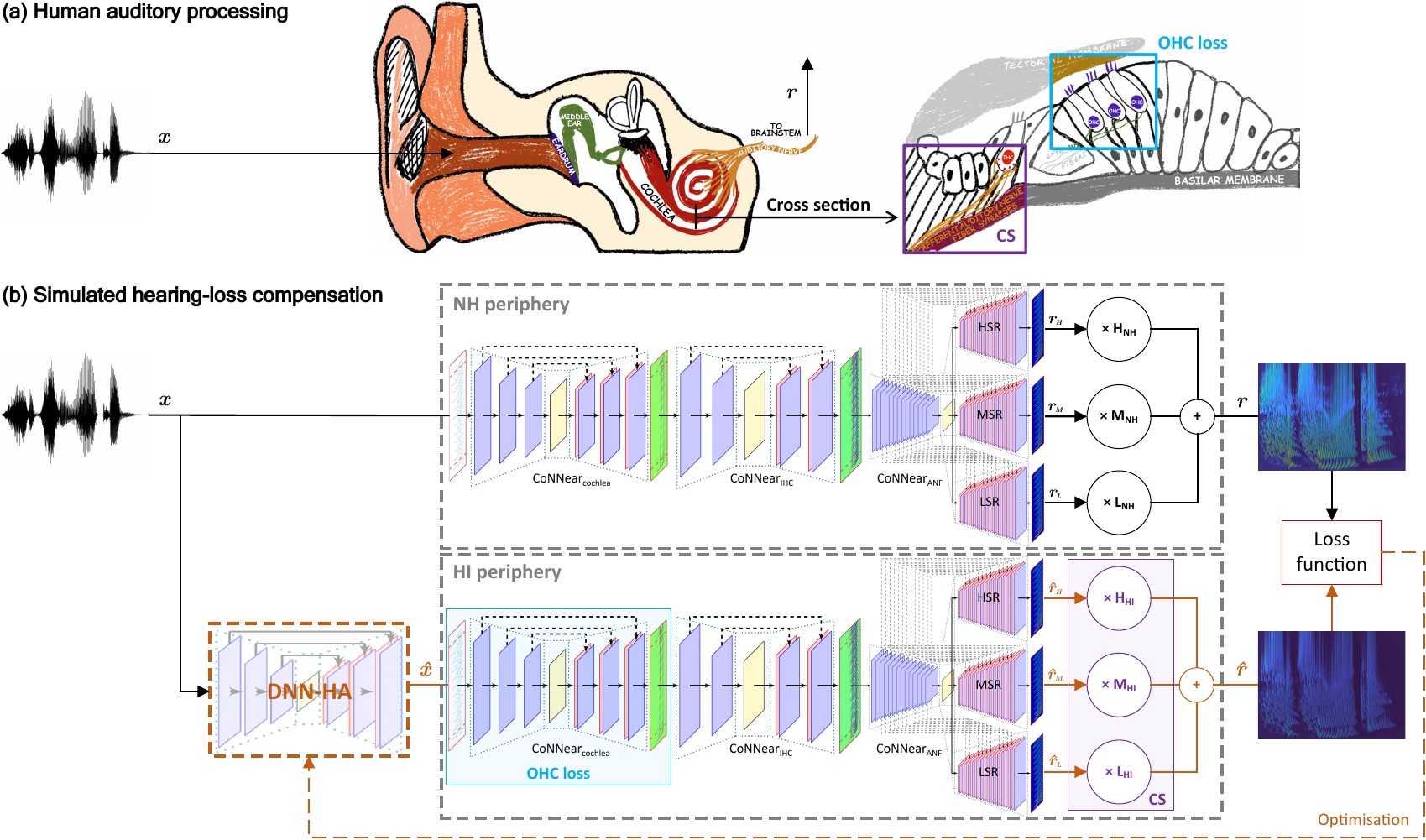}
\end{center}
\vspace{-10pt}
\caption[Framework for hearing-loss compensation.]{Framework for hearing-loss compensation. (a) The human auditory periphery converts an auditory signal $x$ into an electrical signal $r$, transmitted to the brainstem through the auditory nerve. A cross section of the cochlea illustrates the effect of OHC loss and CS in a HI periphery. (b) Diagram of the simulated framework for the design of individualised hearing-aid (HA) models. The DNN-HA model is trained via backpropagation to generate a processed stimulus $\hat{x}$ that minimises the loss function between the NH periphery response $r$ and the HI periphery response $\hat{r}$.}
\label{fig:fw}
\vspace{-10pt}
\end{figure*}

The human auditory system transforms sound into electrical signals via the cochlear mechanics and hair cells, which are then transmitted to the brainstem through the AN.
Figure~\ref{fig:fw}(a) illustrates the human auditory periphery, and the transduction of an auditory signal $x(n)$ via the middle ear and the cochlea to an AN response signal $r(n,w)$, with $n$ and $w$ denoting the time and frequency dimensions, respectively. A cross section through the cochlear turns shows the three major structural components that are responsible for this mechano-electrical transduction: the OHCs, the IHCs, and the afferent ANF synapses. These peripheral components can be impaired by SNHL after ageing or noise exposure \cite{shaheen2015towards, wu2019primary}, leading to fewer/damaged OHCs (OHC loss) or ANF synapse loss (CS) that degrades the transmitted electrical signal $r$.

To optimally compensate for OHC and AN damage in HI auditory systems, we propose a framework for DNN-based HA algorithms (Fig.~\ref{fig:fw}(b)). The framework consists of two pathways: One corresponding to the AN response of the reference NH auditory periphery $r$, and one corresponding to the AN response of a HI periphery $\hat{r}$. 
The derived AN responses $r$ and $\hat{r}$ correspond to biophysically realistic time-frequency representations of sound (neurograms) simulated at distinct cochlear locations, i.e. instantaneous firing rates across different cochlear channels with centre frequencies (CFs) between 112 Hz and 12 kHz \cite{greenwood1990cochlear}.
The HI periphery can be adjusted to reflect the SNHL profile of an individual listener (e.g. based on audiometry or auditory evoked potentials \cite{keshishzadeh2021towards}), by introducing frequency-dependent degrees of OHC loss and/or CS.
Here, we present a numerical optimisation method to determine the parameters of a DNN-HA model that processes an auditory signal such that the HI AN response to the modified signal optimally matches the NH AN response.
In each iteration, the input $x$ is given to the two pathways and the difference between the two generated responses is computed using a pre-defined loss function. Since all components in this framework are differentiable, the DNN-based HA model can be trained via backpropagation to minimise the selected loss function. After training, the DNN-HA model can process an input auditory stimulus $x$ to produce a processed version $\hat{x}$ that, when given as input to the specific HI periphery, will yield an enhanced AN response $\hat{r}$ that optimally matches the reference NH response $r$ (Fig.~\ref{fig:fw}(b)).

\subsection{Auditory periphery models}
Both NH and HI auditory peripheries in Fig.~\ref{fig:fw} are comprised of biophysically inspired CNN-based models that accurately describe human cochlear, IHC and ANF processing: CoNNear\textsubscript{cochlea} \cite{baby2021convolutional} and CoNNear\textsubscript{IHC-ANF} \cite{Drakopoulos2021}. 
As part of our previous studies, we demonstrated that the CoNNear models accurately capture the biophysical properties of an analytical model of the auditory periphery \cite{verhulst_hearres2018} based on an evaluation of key electrophysiological properties described in experimental studies and using data that were not part of the training.
When connected together, these models simulate the responses of three ANF types (high (H), medium (M), and low (L) spontaneous-rate (SR) fibres) to an auditory input signal $x$ across 201 cochlear locations.
In each periphery, the responses of the three ANF types are combined to yield the simulated AN summed responses $r$ and $\hat{r}$ (Fig.~\ref{fig:fw}(b)). 
For the reference NH periphery, the AN summed response $r$ is derived across time and frequency by multiplying the responses of the three ANF types $r_H$, $r_M$ and $r_L$ by the variables $H\textsubscript{NH}$ = 13, $M\textsubscript{NH}$ = 3 and $L\textsubscript{NH}$ = 3 \cite{verhulst_hearres2018}, respectively, that roughly correspond to the distribution of ANF types in cat \cite{liberman1978auditory} and to the total number of ANF innervations to a single IHC \cite{kujawa2009adding}.

\subsubsection{Branched ANF model}
\label{sec:branched}
To capture the slow adaptation properties of the ANF processing stage \cite{westerman_diffusion_1988}, CoNNear\textsubscript{ANF} uses a deep CNN architecture for each ANF type, and this can be computationally expensive for long stimuli \cite{Drakopoulos2021}.
To speed up the training procedure for the present study, we developed a more efficient version of the CoNNear\textsubscript{ANF} model.
The rationale was that a shared encoder would be sufficient to simulate the AN stage, with the processing of each different ANF type (HSR, MSR and LSR fibres) approximated in an individual decoder.
We thus trained a three-branch CNN model with one encoder and three decoders following the procedures outlined in \cite{Drakopoulos2021}.
The hyperparameters of the branched CoNNear\textsubscript{ANF} were identical to the original model \cite{Drakopoulos2021}, with the only difference that a PReLU non-linearity was used in all convolutional layers.
The resulting CoNNear\textsubscript{ANF} architecture is shown in the NH and HI blocks of Fig.~\ref{fig:fw}(b), while the achieved accuracy and execution time are provided in Supplementary Fig.~\ref{fig:rmse_anf} and Supplementary Table~\ref{tab:timing}, respectively.

\subsubsection{Hearing impairment}
\label{sec:hi}
Biophysically inspired computational models of the auditory periphery can accurately describe the different stages of human auditory processing and simulate the impact that different SNHL deficits can have in HI systems \cite{verhulst_hearres2018}.
A unique feature of our framework is that we adopt modular CNN structures for the different peripheral auditory elements such that we can introduce hearing damage in each of these.
Thus, the corresponding CoNNear modules can be adjusted to simulate different degrees of OHC loss and CS, as indicated in the HI block of Fig.~\ref{fig:fw}(b). 
To introduce OHC loss, the NH CoNNear\textsubscript{cochlea} model is retrained using transfer learning based on a specific gain loss profile or an individual audiogram \cite{van2020hearing}. 
Because the biophysical accuracy of each CoNNear periphery component was carefully validated in separate studies \cite{baby2021convolutional, Drakopoulos2021}, HI CoNNear cochlear outputs are expected to yield the same AN responses in CoNNear as in the reference analytical biophysical model \cite{verhulst_hearres2018}. 
To corroborate 
this, we evaluated the outputs of the CoNNear periphery model for two HI profiles (Slope35 and Flat35) that correspond to the highest degree of OHC loss in the model \cite{verhulst_hearres2018},
and show that these HI model responses match the HI responses of the reference analytical model well (Supplementary Fig.~\ref{fig:speech}).

To account for the CS aspect of SNHL, lower values for the variables $H\textsubscript{HI}$, $M\textsubscript{HI}$ and $L\textsubscript{HI}$ can be used (Fig.~\ref{fig:fw}(b)). Previous EEG-based studies have shown that linear decrements of the distribution of ANFs across CF are sufficient to predict individual CS profiles in NH and HI listeners \cite{keshishzadeh2020derived, buran2020predicting, keshishzadeh2021towards, vasilkov2021enhancing}. 
Hence, we adopted single values for the three variables across CF to simulate frequency-independent CS in this study. However, the CS variables can be made frequency specific in the framework if desired.

\subsection{Loss functions} \label{sec:loss}
In our framework, the loss function defines which aspects of the CoNNear model responses will be minimised during training of the DNN-HA model.
Preliminary simulations with the present approach showed that a complete restoration of simulated HI AN responses is not straightforward \cite{drakopoulos2022differentiable}, especially for severe hearing deficits that include CS.
While it might not be possible to fully compensate for HI processing using acoustic treatments in such cases, a partial compensation might still be achieved by constraining the problem and by minimising specific aspects of the simulated AN responses.
To this end, we introduce a variety of additional constraints to the training that focus on the minimisation of features that are known to be functionally relevant for hearing loss and for auditory perception.
The considered loss functions are described here and their performance is evaluated in Section~\ref{sec:results}.

\subsubsection{Time-frequency optimisation}
An ideal hearing-restoration strategy would reduce the difference between the AN responses of the NH and HI peripheries after training to zero:
\begin{equation} \label{eq:lch}
    \ell_{r} = \text{MAE}\{{r(n,w)}, {\hat{r}(n,w)}\},
\end{equation}
where \text{MAE} denotes the mean absolute error, $r$ is the NH AN response to the unprocessed stimulus $x$, $\hat{r}$ is the HI AN response to the processed stimulus $\hat{x}$, and $n$ and $w$ are the time and frequency indices of the AN responses, respectively. 
We chose to use the \text{MAE} in the loss functions, computed as the mean absolute difference (L1-loss) between the two AN responses across time and frequency. 
Theoretically, such a fully unconstrained minimisation would restore the AN response of a HI periphery to that of the NH periphery after training. However, given the highly non-linear character of auditory processing and the corresponding models, along with the severity of introduced SNHL in the HI model, an ideal restoration of the AN responses across time and frequency is expected to be a complex task. 

\subsubsection{Population responses} 
A first constraint was to use the time-domain AN population responses that are obtained by summing the derived AN responses across frequency. 
This summation corresponds to the compound action potential (CAP) and is an important generator source of scalp-recorded auditory-brainstem responses \cite{dau2003importance, bourien2014contribution, verhulst_hearres2018}.
After summing the AN responses across the simulated CFs of the model, the resulting AN population responses (or CAP) was minimised using the mean absolute difference:
\begin{gather}
    \ell_{r_{p}} = \text{MAE}\{r_{p}(n), \hat{r}_{p}(n)\}, \label{eq:lsum} \\  
    r_{p}(n) = \sum_{w=1}^{N_{CF}}{r(n,w)}, \label{eq:rsum} \\ 
    \hat{r}_{p}(n) = \sum_{w=1}^{N_{CF}}{\hat{r}(n,w)},
\end{gather}
where $r_{p}$ and $\hat{r}_{p}$ are the AN population responses, and $N_{CF}$ the number of CFs in the CoNNear models (and in the respective AN responses). 

\subsubsection{Frequency representations} The non-linear nature of the auditory periphery model \cite{verhulst_hearres2018} can result in HI responses that differ in phase compared to the NH responses.
The inclusion of phase-insensitive loss functions could thus help to minimise the magnitude difference of the responses without considering possible time shifts during training.
Thus, we adopted a loss function between the short-time Fourier transform (STFT) magnitudes of the AN responses that was computed separately for each frequency channel:
\begin{gather}
\ell_{R} = \text{MAE}\{R(w,m,k),\hat{R}(w,m,k)\}, \label{eq:lstftch} \\
R(w,m,k) = |\text{STFT}\{r(n,w)\}|,\\ 
\hat{R}(w,m,k) = |\text{STFT}\{\hat{r}(n,w)\}|,
\end{gather}
where $w$ is the frequency index of the AN responses, while $m$ and $k$ are the time and frequency indices of the derived STFT magnitudes $R$ and $\hat{R}$. 
Similarly, a loss function that focusses on the STFT magnitudes of the AN population responses (Eq.~\ref{eq:rsum}) can be computed:
\begin{gather}
\ell_{R_{p}} = \text{MAE}\{R_{p}(m,k),\hat{R}_{p}(m,k)\},  \label{eq:lstftsum}\\
R_{p}(m,k) = |\text{STFT}\{r_{p}(n)\}|, \\ 
\hat{R}_{p}(m,k) = |\text{STFT}\{\hat{r}_{p}(n)\}|, \label{eq:rhstftsum}
\end{gather}

\subsubsection{Contrast enhancement} 
Many studies have shown that temporal-envelope coding is essential for robust speech intelligibility and that it is degraded by CS \cite{fullgrabe2009contribution, kubanek2013tracking, souza2021does, parthasarathy2019age, vasilkov2021enhancing}, and our previous study \cite{drakopoulos2022model} showed that emphasising the temporal peaks of sound can lead to increased physiological and behavioural responses.
To address this and target the enhancement of the temporal modulations of speech, we computed the loss functions of Eqs.~\ref{eq:lch} and \ref{eq:lsum} also for the squared NH and HI AN responses ($r^2$ and $\hat{r}^2$).
This focusses the optimisation on the most stimulated regions and temporal peaks of the responses, and less on troughs or silent parts in the audio.
Similarly, the spectral contrast can also be emphasised by squaring the STFT magnitudes in the loss functions of Eqs.~\ref{eq:lstftch} and \ref{eq:lstftsum} (power spectrograms).

\subsubsection{Frequency limitation} Although the CoNNear models operate at a sampling frequency $f_{s}$ = 20 kHz, the adopted training dataset \cite{garofolo1993darpa} had a sampling frequency of 16 kHz. To ensure that AN responses can be enhanced by processing sound at frequencies below 8 kHz (Nyquist frequency of the dataset), we added a loss function that minimises the magnitude differences between the original and processed stimuli ($x$ and $\hat{x}$) at frequencies above 8 kHz:
\begin{gather}
\ell_{X} = \text{MAE}\{X(k),\hat{X}(k)\} \text{ for } k > 8 \text{ kHz}, \\
X(k) = |\text{FFT}\{x(n)\}|, \\
\hat{X}(k) = |\text{FFT}\{\hat{x}(n)\}|,
\end{gather}
where FFT denotes the fast Fourier transform and $k$ the frequency index of the derived FFT spectra. 

\section{Training} \label{sec:training}
To train our DNN-HA models, we used 2310 randomly selected recordings from the TIMIT speech corpus \cite{garofolo1993darpa} that we upsampled to 20 kHz and calibrated to 70 dB sound pressure level (SPL) using a reference of $p_{0}=2\cdot10^{-5}$~Pa. 
To account for the required context window of the CoNNear ANF model \cite{Drakopoulos2021}, $L\textsubscript{l}$ = 7,936 and $L\textsubscript{r}$ = 256 samples of silence (zeros) were added at the beginning and end of each sentence, respectively.
Additionally, CoNNear\textsubscript{ANF} requires inputs that are multiples of 16,384 samples, thus each sentence was zero-padded at the end to yield a training dataset with total size of $L\textsubscript{c}$ = 81,920 samples. 
No noise was added to the dataset, to ensure that the trained HA processing does not focus on noise suppression to enhance the degraded AN responses, but rather focusses on an optimal stimulation of ANFs.
Future developments can of course target the restoration of distorted speech (e.g. noisy or reverberated) as well.

By default, the CoNNear models simulate responses across $201$ tonotopic cochlear locations \cite{baby2021convolutional}, corresponding to CFs between 112 Hz and 12 kHz spaced according to the Greenwood place-frequency map of the cochlea \cite{greenwood1990cochlear}. Here, $N\textsubscript{CF}$ = 21 frequency channels were selected out of the 201 (channels 1 to 201 with a step of 10) to simplify and speed up the training procedure.
Thus, for inputs $x$ of $L\textsubscript{c}$ = 81,920 samples, the CoNNear models generated AN responses $r$ with size $L \times N_{\text{CF}}$, where $L$ = 73,728 samples corresponds to the time-dimension length after accounting for the context of CoNNear\textsubscript{ANF} \cite{Drakopoulos2021}. 
The layer weights of all CoNNear models were fixed during training, thus allowing only the parameters of the DNN-HA model to be optimised. 

\subsection{DNN-HA models}
The DNN-HA models used an end-to-end, encoder-decoder CNN architecture \cite{baby2019sergan, drakopoulos_ica2019} with 16 convolutional layers (i.e. 8 in the encoder and 8 in the decoder) and [16, 32, 32, 64, 64, 128, 128, 256] filters in each encoder layer. These were mirrored in reverse order in the decoder layers, while each layer had a filter length of 32 and was followed by a PReLU non-linearity (except for the last layer). 
This resulted in a CNN architecture with 5,197,633 trainable parameters.
A stride of 2 was used in each convolutional encoder layer to halve the temporal (time) dimension of the input, resulting in a shrunk dimension of $L/2^8$ samples at the end of the encoder.
Skip connections were added to each layer and the decoder followed the opposite procedure to double the temporal dimension and generate an output with the same size as the input signal (end-to-end processing).
The chosen architecture was earlier shown to be efficient for speech enhancement applications \cite{drakopoulos_ica2019, baby2019sergan}.

During training, the context before ($L\textsubscript{l}$) and after ($L\textsubscript{r}$) the input sentences was cropped out and the resulting cropped inputs ($L$ = 73,728 samples) were processed by the DNN-HA model, to avoid a compensation for the context parts which were not considered in the simulated AN responses and loss functions.
After processing, the cropped-out context of the original input was concatenated to the processed signal and the total stimulus ($L\textsubscript{c}$ = 81,920 samples) was given to the CoNNear models.
Furthermore, the cropped signals were segmented into sub-windows of $L\textsubscript{p}$ = 2048 samples (102.4 ms for $f\textsubscript{s}$ = 20 kHz) before they were given as inputs to the DNN-HA models. 
This step ensures that the same processing will be applied to the input signal regardless of the adaptation properties of the AN responses \cite{Drakopoulos2021}, while also rendering the trained DNN-HA models suitable for real-time applications.
Although the DNN-HA models were trained to process sub-frames of $L\textsubscript{p}$ = 2048 samples, after training, they can process inputs of any size (multiples of $2^8 = 256$ samples for 8 encoder layers).

\begin{table}[t!]
  \begin{center}
  \caption[Trained DNN-HA models.]{Eight DNN-HA models were trained using different combinations of the loss functions of Section~\ref{sec:loss}, thus minimising different aspects of the simulated AN responses. The square symbol on the losses ($^2$) indicates that squared AN representations were used in the respective loss functions (squared time-domain responses or squared STFT magnitudes).}
\vspace{-5pt}
  \aboverulesep=0ex
  \belowrulesep=0ex
  \begin{tabular}{l | l}
  \toprule
  Name & Joint loss function \\ 
  \midrule
  $\La_{r}$ & $1 \cdot \ell_{r}$, $0.5 \cdot \ell_{X}$ \\
  $\La_{r,R}$ & $1 \cdot \ell_{r}$, $0.5 \cdot \ell_{X}$, $0.1 \cdot \ell_{R}$ \\
  $\La_{r,r_p}$ & $1 \cdot \ell_{r}$, $0.5 \cdot \ell_{X}$, $0.1 \cdot \ell_{r_{p}}$ \\
  $\La_{r,r_p,R_p}$ & $1 \cdot \ell_{r}$, $0.5 \cdot \ell_{X}$, $0.1 \cdot \ell_{r_{p}}$, $0.02 \cdot \ell_{R_{p}}$ \\
  $\La_{r^2}$ & $1 \cdot \ell_{r^{2}}$, $40 \cdot \ell_{X}$ \\
  $\La_{r^2,R^2}$ & $1 \cdot \ell_{r^{2}}$, $40 \cdot \ell_{X}$, $0.0014\cdot \ell_{R^{2}}$ \\ 
  $\La_{r^2,{r^2_p}}$ & $1 \cdot \ell_{r^{2}}$, $40 \cdot \ell_{X}$, $0.08 \cdot \ell_{r^2_p}$ \\
  $\La_{r^2,{r^2_p},{R^2_p}}$ & $1 \cdot \ell_{r^{2}}$, $40 \cdot \ell_{X}$, $0.08 \cdot \ell_{r^2_p}$, $10^{-5}\cdot \ell_{R^2_p}$ \\
  \bottomrule
  \end{tabular}
  \label{tab:models}
	\end{center}
\vspace{-15pt}
\end{table}

To compare the outcomes of different training setups, eight DNN-HA models were trained using combinations of the loss functions introduced in Section~\ref{sec:loss} (Table~\ref{tab:models}).
Different weighting factors were used to ensure that all individual losses had similar contributions during training.
Although different aspects were minimised each time (population responses, STFT magnitudes, squared responses), the $\ell_r$ loss function (Eq.~\ref{eq:lch}) was used in all models (before or after squaring) to ensure an equally distributed enhancement across frequency.
For the computation of the STFT (Eqs.~\ref{eq:lstftch}-\ref{eq:rhstftsum}), Hanning windows of $L\textsubscript{p}$ = 2048 samples were used with 50\% overlap.
The entire framework was developed using the Tensorflow \cite{abadi2016tensorflow} and Keras \cite{chollet2015keras} machine-learning libraries. 
Training was performed for 60 epochs, using a batch size of 1 for the training dataset and an Adam optimiser \cite{kingma2014adam}. A sampling frequency of 20 kHz was used for the CoNNear and DNN-HA models.

\section{Evaluation}

To assess the restoration capabilities of our framework, we trained the DNN-HA models to compensate for a HI profile that comprised: A sloping high-frequency audiogram (OHC loss) starting from 1 kHz with elevated hearing thresholds of 35 dB HL at 8 kHz, abbreviated as \emph{Slope35} (inset of Fig.~\ref{fig:ANsummed}(b)), and a complete loss of LSR and MSR ANFs ($L\textsubscript{HI}$ = 0 and $M\textsubscript{HI}$ = 0) and a $\sim$46\% loss of HSR ANFs ($H\textsubscript{HI}$ = 7), abbreviated as \emph{7,0,0}.
This combination of sensorineural hearing deficits reflects the average predicted hearing profile among older NH people ($64.25\pm1.88$ years) from our previous study \cite{keshishzadeh2021towards}.
Data from post-mortem temporal-bone studies demonstrate a 50\% AN innervation loss after the age of 50 \cite{viana2015cochlear, wu2019primary}, motivating our choice for such a severe ANF damage profile.

To quantify the benefit of each trained model on HI sound processing, the Flemish Matrix corpus \cite{luts2014development} was used. This test dataset includes 260 sentences recorded by a female speaker, each comprising a 5-word combination of a closed set of 50 Dutch words.
The Flemish Matrix material was calibrated to 70 dB SPL, and 
context of 7,936 and 256 samples was added at the beginning and end of each sentence, respectively, when simulating the CoNNear AN responses.
Even though we used $N\textsubscript{CF}$ = 21 frequency channels to speed up and simplify the training procedure of the DNN-HA models, the full 201-channel CoNNear models were used for the evaluation. Thus, all following figures and tables correspond to simulated responses of the full-channel models. 

\subsection{Simulated DNN-HA processing benefits in quiet} \label{sec:sim_clean}

We first evaluated the trained DNN-HA models based on the restoration of the simulated HI AN responses.
Restoration success was quantified using the difference between the simulated NH AN population response to an unprocessed stimulus and the simulated HI AN response to the DNN-HA processed stimulus.
To this end, the average RMSE was computed between the simulated NH and HI AN population responses across the 260 sentences of the Flemish Matrix, normalised to the maximum of the NH response for each sentence:
    \begin{gather} 
\text{NRMSE} = \frac{\text{RMSE}}{\max{(r_{p})}}, \\ \label{eq:nrmse}
\text{RMSE} = \sqrt{\frac{\sum_{n=1}^{N}(r_{p}(n)-\hat{r}_{p}(n))^2}{N}},
    \end{gather}
where $r_{p}$ and $\hat{r}_{p}$ are the NH and HI AN population responses (Eq.~\ref{eq:rsum}) respectively, $n$ corresponds to each sample of the AN population responses, and $N$ to the total number of samples. While the computed RMSE has the same measurement unit as the estimated quantity (spikes/s), the normalised errors (NRMSEs) are expressed in percentages (\%) of the NH response maximum. 
To evaluate the algorithms for conversational speech, the NRMSEs were computed for sentences with root-mean-square (RMS) energy levels between 30 and 70 dB SPL using a step of 10 dB.

Since recent studies have shown that CS degrades auditory processing of temporal-envelope information in sound \cite{parthasarathy2019age,vasilkov2021enhancing}, we also took this aspect into account when evaluating restoration success for CS-compensating DNN-HA models.
A metric that was earlier used to evaluate auditory temporal-envelope processing is the enhancement of population AN responses to amplitude-modulated stimuli \cite{keshishzadeh2020derived, vasilkov2021enhancing, drakopoulos2022model}.
To calculate the benefit after DNN-HA processing, we used a 400-ms, 70-dB-SPL sinusoidally amplitude-modulated (SAM) tonal stimulus (carrier frequency $f_{c} = 4$ kHz, modulation frequency $f_{m} = 120$ Hz and $m = 100\%$ modulation \cite{vasilkov2021enhancing, drakopoulos2022model}) with a Hanning window of 5~ms.
The cochlear-nucleus and inferior-colliculus modules of the analytical periphery model \cite{verhulst_hearres2018} were used as a backend to the CoNNear models to derive the envelope-following responses (EFRs) from the AN population responses, which were then used to compute the \text{EFR\textsubscript{sum}} metric:
  \begin{equation} \label{eq:EFRs}
\text{EFR\textsubscript{sum}} = \sum_{i=0}^{3}{M_{f_{i}}},
  \end{equation}
where $M_{f_{i}}$ are the spectral peaks of the simulated EFR at the modulation frequency $f_{0} = f_{m}$ of the stimulus and the next 3 harmonics of $f_{0}$ \cite{keshishzadeh2021towards, drakopoulos2022model}. For the NH periphery, this resulted in an EFR\textsubscript{sum} magnitude of 8.39 nV (reference). 

\subsection{Simulated DNN-HA processing benefits in noise}

It is important to test DNN-HA performance also for noisy speech, as general communication is not limited to quiet places only.
To this end, we evaluated speech processing and perception in noise for our DNN-HA models using well-established objective metrics.
At first, we selected three SNR conditions of -12, -6 and 0 dB. Experimental results of previous studies that used the Flemish Matrix test showed $\sim$15\% word recognition at -12 dB SNR and $\sim$85\% at -6 dB SNR for young NH subjects \cite{maele2021variability, drakopoulos2022model}, thus these two SNR conditions are expected to roughly correspond to scenarios of challenging and easy speech recognition in noise, respectively. Speech-shaped noise (SSN) was generated from the Flemish Matrix material and added to the evaluation sentences at the desired SNR level.

\begin{table}[t!]
  \begin{center}
  \caption[Simulated DNN-HA processing benefits in noise.]{Nine objective metrics were considered to evaluate DNN-HA processing benefits for speech in noise. For each metric, we report the implementation that was used, the reference signal of the prediction (if any), and the sampling frequency $f_s$.}
\vspace{-5pt}
  \aboverulesep=0ex
  \belowrulesep=0ex
  \begin{tabular}{l | r r r}
  \toprule
  Metric & Implementation & Reference & $f_s$ (kHz) \\ 
  \midrule
  PESQ \cite{rix2001perceptual} & Python\tablefootnote{\url{https://github.com/ludlows/python-pesq}} & clean signal & 16 \\
  STOI \cite{taal2011algorithm}, ESTOI \cite{jensen2016algorithm} & Python\tablefootnote{\url{https://github.com/mpariente/pystoi}} & clean signal & 20 \\
  SIIB \cite{van2017instrumental} & Python\tablefootnote{\url{https://github.com/kamo-naoyuki/pySIIB}} & clean signal & 20 \\
  SRMR \cite{falk2010non, santos2014improved} & Python\tablefootnote{\url{https://github.com/jfsantos/SRMRpy}} & - & 20 \\
  mr-sEPSM \cite{jorgensen2013multi} & MATLAB \cite{majdak2021amt} & noise signal & 44.1 \\
  sEPSM\textsuperscript{corr} \cite{relano2016predicting} & MATLAB \cite{majdak2021amt} & clean signal & 44.1 \\
  sCASP \cite{relano2019speech} & MATLAB \cite{majdak2021amt} & clean signal & 44.1 \\
  HASPI v2 \cite{kates2021hearing} & MATLAB \cite{kates2021hearing} & clean signal & 44.1 \\
  FADE \cite{schadler2016simulation} & MATLAB \cite{schadler2016simulation} & - & 44.1 \\
  \bottomrule
  \end{tabular}
  \label{tab:metrics}
	\end{center}
\vspace{-15pt}
\end{table}

Nine different objective metrics were considered from the literature: One for speech quality prediction (PESQ) and the remaining eight for speech intelligibility.
Although the existing objective metrics are not explicitly designed for the type of non-linear processing that our DNN-based HA strategies generate and might therefore not generalise to accurately predict the performance of NH and HI listeners, this evaluation step was used to provide a rough indication of the perceptual effects that such processing strategies can have for speech in noise.
A number of different metrics were thus included in our evaluation stage to ensure that any observed benefits are shared across multiple metrics.
The metrics are summarised in Table~\ref{tab:metrics}, and more details can be found in Appendix A.
Similar to how we evaluated the DNN-HA models for speech in quiet, we also computed the average NRMSE between the NH AN population responses to the unprocessed noisy sentences and the HI AN population responses to the DNN-HA processed noisy sentences.

\section{Results} \label{sec:results}

\begin{table}[t!]
  \begin{center}
  \caption[Simulated CoNNear evaluation in quiet.]{Simulated CoNNear evaluation in quiet. The average NRMSEs and EFR magnitudes of the HI periphery (Slope35-7,0,0) were computed before and after processing with each trained DNN-HA model. For levels from 30 to 70 dB SPL, the average NRMSEs were computed between the NH AN population responses to the unprocessed sentences and the HI AN population responses to the unprocessed and processed sentences of the Flemish Matrix. The EFR magnitudes were computed from the simulated EFRs of the HI AN population responses to a SAM tonal stimulus (Eq.~\ref{eq:EFRs}). The EFR reference magnitude was 8.39 nV (NH). In each case, the loss function that gave the best result is indicated in bold.}
\vspace{-10pt}
  \setlength{\tabcolsep}{5.5pt}
  \aboverulesep=0ex
  \belowrulesep=0ex
  \begin{tabular}{l | c c c c c | c}
  \toprule
  & \multicolumn{5}{c | }{NRMSE (\%)} & EFR\textsubscript{sum} (nV) \\ 
  & 30 & 40 & 50 & 60 & 70 & 70 (dB SPL) \\ 
  \midrule
  Unprocessed & 40.06 & 28.41 & 19.79 & 15.40 & 13.78 & 4.31 \\
  \hline
  $\La_{r}$ & 26.47 & 17.97 & 13.93 & 12.60 & 12.50 & 3.61 \\
  $\La_{r,R}$ & 29.04 & 20.14 & 14.73 & 12.86 & 12.74 & 4.17 \\
  $\La_{r,r_p}$ & \textbf{10.04} & \textbf{8.42} & \textbf{8.32} & \textbf{10.27} & 12.15 & 3.25 \\
  $\La_{r,r_p,R_p}$ & 13.32 & 10.60 & 9.67 & 10.66 & 11.90 & 6.12 \\
  $\La_{r^2}$ & 30.40 & 23.90 & 17.15 & 14.09 & 13.24 & 5.96 \\
  $\La_{r^2,R^2}$ & 22.87 & 18.86 & 15.14 & 13.32 & 12.99 & 5.36 \\
  $\La_{r^2,r^2_p}$ & 26.20 & 20.90 & 15.28 & 13.04 & 12.89 & 6.07 \\
  $\La_{r^2,r^2_p,R^2_p}$ & 19.18 & 15.82 & 11.58 & 10.41 & \textbf{11.81} & \textbf{7.75} \\
  \bottomrule
  \end{tabular}
  \label{tab:clean}
	\end{center}
\vspace{-15pt}
\end{table}

\begin{figure*}[tb!]
\begin{center}
\includegraphics[width=\textwidth,height=\textheight,keepaspectratio]{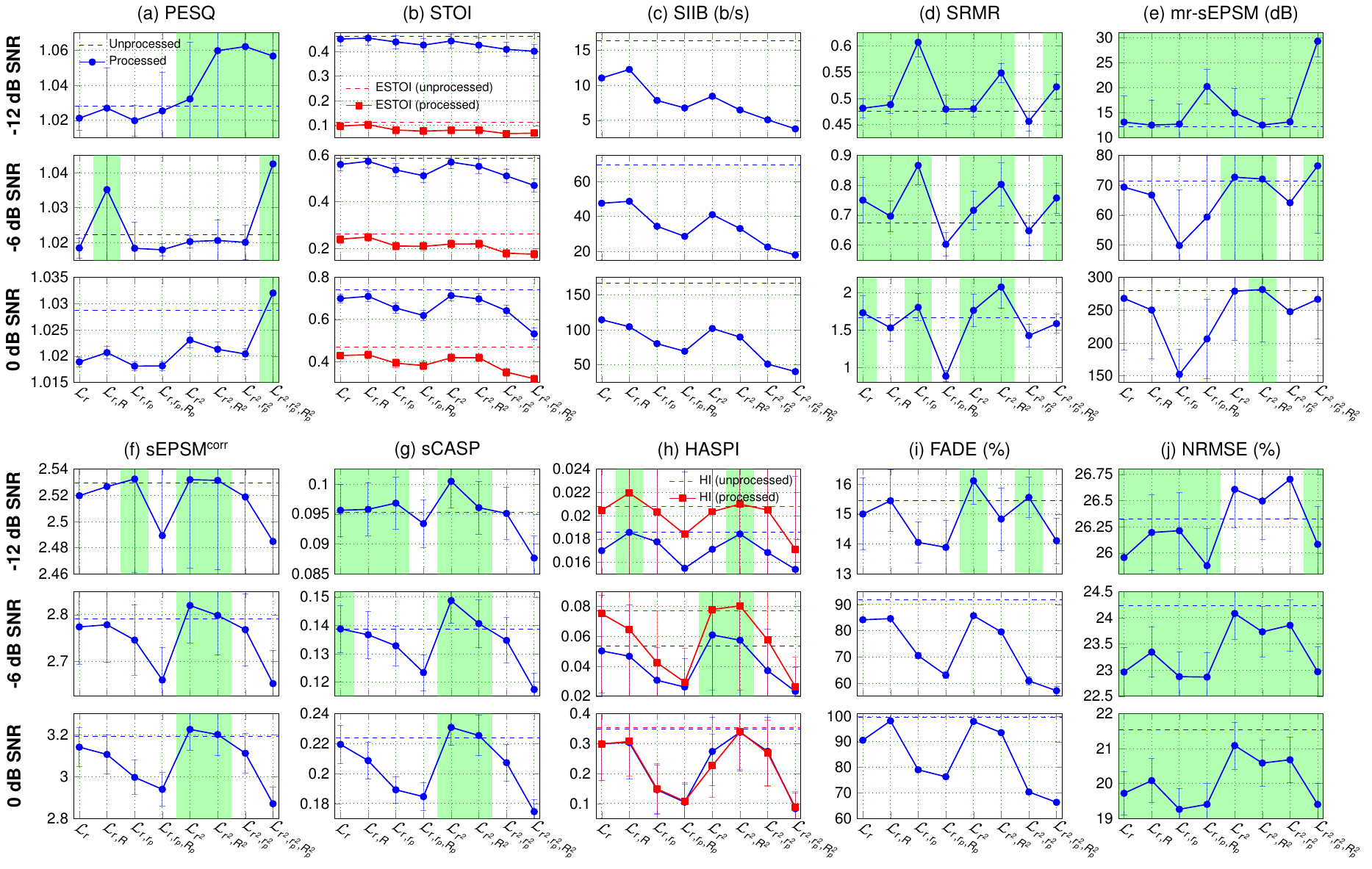}
\end{center}
\vspace{-15pt}
\caption[Objective evaluation for speech in noise (Slope35-7,0,0).]{Objective evaluation for speech in noise (Slope35-7,0,0). The trained DNN-HA models were evaluated using different objective metrics (cf. Table~\ref{tab:metrics}), computed for the 70-dB-SPL Flemish Matrix sentences in noise at 3 SNRs (-12, -6 and 0 dB). For each metric at each SNR condition, the dashed line corresponds to the average score of the unprocessed sentences, and the vertical bars to the standard deviation of the scores among the different processed sentences. The green-hued boxes indicate loss functions for which a better average performance was achieved after processing. Panel (j) shows the same results for the NRMSEs of the NH and HI AN population responses.}
\label{fig:scores}
\vspace{-10pt}
\end{figure*}

The performance of the eight trained DNN-HA models (Table~\ref{tab:models}) in quiet and in noise is evaluated in Table~\ref{tab:clean} and Figs.~\ref{fig:ANsummed}-\ref{fig:optim}. The best-performing models were further improved by applying additional constraints to their training losses (Section \ref{sec:simpl}).
The optimal training setup was determined and applied to two less severe HI profiles to test generalisation of our methods to other SNHL profiles (Figs.~\ref{fig:ANs_simpl}-\ref{fig:simpl}).

\subsection{DNN-HA processing benefits in quiet}

Table~\ref{tab:clean} shows the average NRMSEs between the NH and HI AN population responses, computed for the Flemish Matrix sentences before and after processing.
The NRMSE results are expressed as difference percentages, with all the discussed improvements corresponding to absolute NRMSE differences (expressed in \% as well).
All eight DNN-HA models reduced the NRMSEs of the AN population responses, but the loss function which focussed on a time-domain minimisation ($\La_{r,r_p}$) showed the best NRMSE benefit by decreasing the difference with percentages between $\sim$30\% (30 dB SPL) and $\sim$1.6\% (70 dB SPL). 
Overall, the processing was more effective for speech presented at lower levels, resulting in HI responses that, on average, differed by $\sim$9\% from the NH responses for speech between 30 and 50 dB SPL (compared to $\sim$29\% before processing). 
The degree of CS in the chosen HI profile was expected to have a stronger effect on the firing rate of the ANFs at low intensity levels (unprocessed NRMSEs decreasing from 40\% to 13.8\% in Table~\ref{tab:clean}), given that only half of the HSR fibres are available in the HI system to encode sound at low levels. This suggests that a partial compensation can be achieved for such HI profiles by simply amplifying signals presented at low levels.
It should also be noted that the NRMSE differences between the NH and HI responses were large across conversational levels for the selected HI profile (up to 40\% of the maximum AN firing rate), suggesting that a complete restoration of HI profiles that include CS might not be possible using audio-processing treatments alone.
This goes in line with our previous work, where we found that the CS aspect of SNHL dominated the difference between the NH and HI AN responses \cite{drakopoulos2022differentiable}.

\begin{figure}[tb!]
\begin{center}
\includegraphics[width=0.49\textwidth,height=\textheight,keepaspectratio]{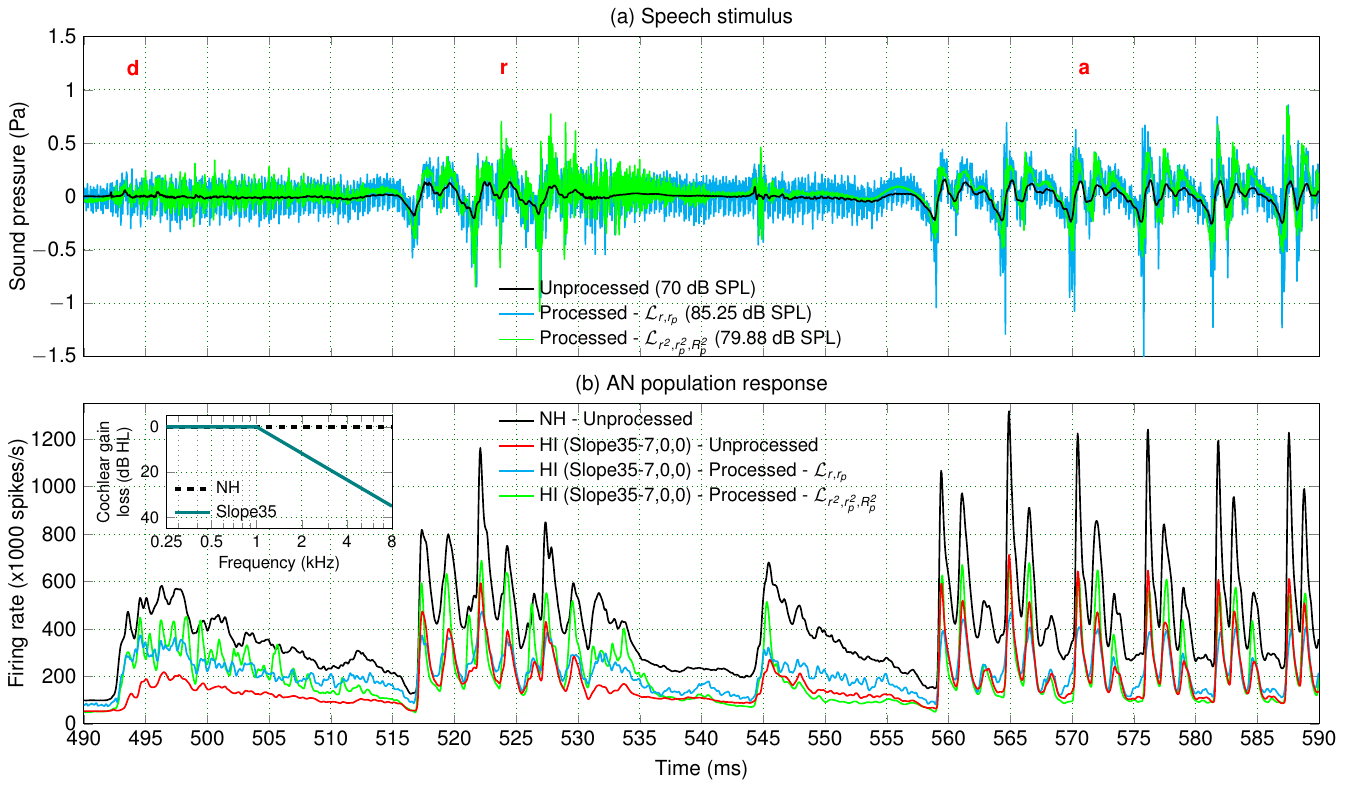}
\end{center}
\vspace{-10pt}
\caption[Simulated AN population responses for two trained DNN-HA models.]{Simulated AN population responses for two trained DNN-HA models. (a) A clean speech segment is shown before and after processing with models $\La_{r,r_p}$ and $\La_{r^2,r^2_p,R^2_p}$, containing two consonants (\textbackslash d\textbackslash , \textbackslash r\textbackslash) and a vowel (\textbackslash a\textbackslash ) from the Flemish word `draagt' (extracted from the sentence `David draagt drie gele boten'). Both models were trained to restore a HI periphery that included both OHC loss and CS (Slope35-7,0,0). (b) Different aspects of the HI AN population response (red) were enhanced after processing the input stimulus with models $\La_{r,r_p}$ (cyan) and $\La_{r^2,r^2_p,R^2_p}$ (green). Ideal restoration would match the respective simulated HI response after processing to the NH response (black).}
\label{fig:ANsummed}
\vspace{-10pt}
\end{figure}

Table~\ref{tab:clean} also shows the simulated EFR magnitudes as an objective marker of peripheral envelope coding (Eq.~\ref{eq:EFRs}), which showed a different trend across DNN-HA models: The first three models failed to improve the EFR\textsubscript{sum} metric, and all remaining models enhanced the simulated magnitudes by more than 1 nV. Loss function $\La_{r^2,r^2_p,R^2_p}$ showed the best EFR benefit by restoring almost completely the EFR amplitude to the NH amplitude (8.39~nV). The last four models ($\La_{r^2}$-$\La_{r^2,r^2_p,R^2_p}$) focussed on the enhancement of the response peaks, and models $\La_{r^2,R^2}$ and $\La_{r^2,r^2_p,R^2_p}$ additionally included the frequency representations of the AN responses, which explains their superior enhancement on the spectral AN response peaks and thus on the EFRs.
It should be noted that the NRMSE metric is sensitive to response regions that are not enhanced by the processing (outliers), and this is reflected by the larger errors obtained for these models (trained on the squared time-domain responses).

To better understand which sound and AN-response features the different loss functions focus on, Fig.~\ref{fig:ANsummed} visualises time-domain sound waveforms and AN responses before and after processing. 
A speech segment was used as input to the NH and HI CoNNear models to simulate the unprocessed AN population responses, and was also processed by two DNN-HA models (Fig.~\ref{fig:ANsummed}(a)) and used as input to the HI model to simulate the AN population responses after processing.
Models $\La_{r,r_p}$ and $\La_{r^2,r^2_p,R^2_p}$ were selected because they yielded the best NRMSE and EFR\textsubscript{sum} scores, respectively (Table~\ref{tab:clean}). 
Figure~\ref{fig:ANsummed}(a) shows that model $\La_{r,r_p}$ introduced high-frequency fluctuations to the stimulus after processing, which enhanced the AN population responses everywhere but not fully restored them to the reference NH responses (Fig.~\ref{fig:ANsummed}(b)). 
Model $\La_{r^2,r^2_p,R^2_p}$ targeted the restoration of the most excited regions of the AN responses, and resulted in a better enhancement of the temporal peaks of the response while only minimally enhancing (or deteriorating in some cases) the response troughs.
Thus, these two ways of semi-constrained training resulted in different compensation strategies that are expected to differ in sound perception or quality as well, as the first selected model ($\La_{r,r_p}$) was trained to minimise time-domain AN responses, while the second model ($\La_{r^2,r^2_p,R^2_p}$) was trained to minimise squared time-domain AN responses and power spectrograms of the AN population responses (Table~\ref{tab:models}).
The enhancement differences of the two loss functions are also visualised in Supplementary Figs.~\ref{fig:AN_SAM} and \ref{fig:AN}, showing time-domain and neurogram responses to the SAM tone stimulus and the speech segment. 

\subsection{DNN-HA processing benefits in noise}
The different panels in Fig.~\ref{fig:scores} show the performance of the selected objective metrics (Table~\ref{tab:metrics}) for speech in noise presented at 3 SNRs. The metric scores were computed for each of the eight trained DNN-HA models and are compared to the unprocessed results (dashed horizontal lines), with the green-hued boxes indicating the DNN-HA models that yielded better performance than the unprocessed reference.
Overall, there were few cases where the scores improved after processing at 0 dB SNR, but at negative SNRs, the PESQ, SRMR, mr-sEPSM, sEPSM\textsuperscript{corr}, sCASP and HASPI metrics showed better performance after processing, especially for models $\La_{r^2}$ and $\La_{r^2,R^2}$.

Additionally, Fig.~\ref{fig:scores}(j) showed improved NRMSEs between the NH and HI AN population responses in most cases after processing, with lower NRMSEs corresponding to better improvements. Models $\La_{r,r_p,R_p}$ and $\La_{r^2,r^2_p,R^2_p}$ showed the greatest benefit (NRMSE decrease of $1.3$\% on average), while small RMSE benefits were found for model $\La_{r^2}$ (up to $\sim$0.5\%).
Thus, different conclusions can be drawn about the best loss function depending on the selected metric. The NRMSE metric showed best results for loss functions that included the AN population responses (e.g. $\La_{r^2,r^2_p,R^2_p}$), while most intelligibility metrics showed benefits for loss functions that minimised the squared AN responses (e.g. $\La_{r^2,R^2}$).

\subsection{Determining the optimal training procedure}
\label{sec:optim}

Based on our speech-in-noise assessment, we selected the best-performing models and included additional constraints to their loss functions, with the aim to further optimise and improve their restoration capabilities. 
To this end, loss functions $\La_{r}$, $\La_{r,R}$, $\La_{r^2}$, $\La_{r^2,R^2}$ and $\La_{r^2,r^2_p,R^2_p}$ were chosen since they yielded improved scores for most of the metrics in the lowest SNR conditions (Fig.~\ref{fig:scores}). 
For this additional analysis, we only considered the two noisiest SNR conditions (-12 and -6 dB SNR) and four objective metrics, as the others did not reflect a consistent benefit after processing.
This resulted in the following: 
The SRMR blind metric (Fig.~\ref{fig:scores}(d)), the sCASP model (Fig.~\ref{fig:scores}(g)) and the HASPI model with the HL profile included (Fig.~\ref{fig:scores}(h)). The results of the ESTOI metric and the FADE model were also computed as references.
Lastly, the NRMSE was used alongside the objective metrics to reflect the benefit in simulated response restoration (Fig.~\ref{fig:scores}(j)). 
All objective metrics were normalised to ranges from 0 to 100 such that all scores can be expressed as a percentage of their maximum values. The maximum value for the SRMR metric was set as the average score of the metric on the clean unprocessed sentences (13.2 metric units). The NRMSE results were inverted by subtracting them from their maximum value (NRMSE' = 1 - NRMSE), such that higher values (i.e. lower NRMSEs) reflect improvement.

\begin{figure*}[tb!]
\begin{center}
\includegraphics[width=0.95\textwidth,height=\textheight,keepaspectratio]{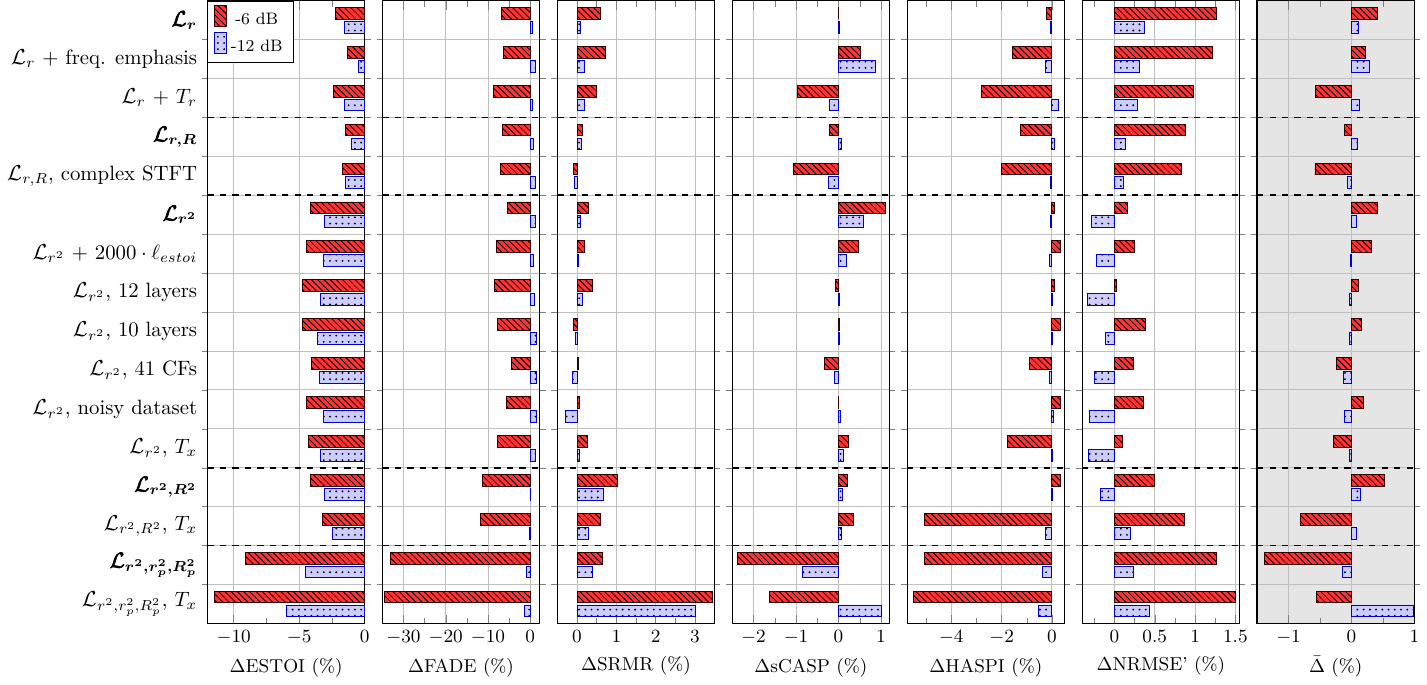}
\end{center}
\vspace{-10pt}
\caption[Improving speech-in-noise performance (Slope35-7,0,0).]{Improving speech-in-noise performance (Slope35-7,0,0). Additional constraints were added to the best-performing loss functions to improve speech intelligibility in noise. Six of the metrics were computed for the 70-dB-SPL Flemish Matrix sentences in noise at 2 SNRs (-6 and -12 dB). The metric scores were normalised to their maximum values and are given as a percent (\%) improvement from the respective unprocessed metric scores, computed by subtracting each reference normalised score (before processing) from the normalised scores after processing. The last metric ($\bar{\Delta}$) corresponds to the overall percent improvement after processing, computed from the SRMR, sCASP, HASPI and NRMSE scores using Eq.~\ref{eq:Dm}. Positive values correspond to improved speech intelligibility after processing.}
\label{fig:optim}
\vspace{-10pt}
\end{figure*}

The different panels in Fig.~\ref{fig:optim} show the achieved processing benefit (in \% improvement) for the different objective metrics and the five selected models (in bold).
The average improvement $\bar{\Delta}$ is also reported in the rightmost column of Fig.~\ref{fig:optim} and was computed from the mean of the normalised score improvement for the SRMR, sCASP, HASPI and NRMSE metrics:
  \begin{equation} \label{eq:Dm}
\bar{\Delta} = \frac{\Delta \text{SRMR} + \Delta \text{sCASP} + \Delta \text{HASPI} + \Delta \text{NRMSE'}}{4},
  \end{equation}
expressed as a percentage improvement (\%) over the respective unprocessed scores. Supplementary Table~\ref{tab:optim_all} reports the original scores for all considered objective metrics.

Overall, Fig.~\ref{fig:optim} shows that the five selected DNN-HA models (in bold) achieved comparable improvements to the six computed metrics, with small benefits on the average score improvement $\bar{\Delta}$ (up to $\sim$0.5\%).
Depending on the individual components included in the loss functions (Table~\ref{tab:models}), different constraints were added to further improve the minimisation procedure and the trained aspects of each DNN-HA model. The different training setups are explained in Appendix B, with the results of the newly trained models juxtaposed with the original models in Fig.~\ref{fig:optim}.

Although objective speech-in-noise improvements were observed for the considered Slope35-7,0,0 HI profile after processing (Fig.~\ref{fig:optim}), the DNN-HA models were not able to surpass a 4\% benefit in any of the computed metrics. Overall, the inclusion of a stimulus threshold $T_{x}$ in loss function $\La_{r^2,r^2_p,R^2_p}$ ($\La_{r^2,r^2_p,R^2_p}$, $T_{x}$; see Appendix B for more details) yielded the best $\bar{\Delta}$ score ($\sim$1\% at -12 dB SNR) and the highest SRMR and NRMSE improvement across all models ($>$3\% improvement on the SRMR over the unprocessed scores).
This trained model decreased the RMSEs by 0.4\% at -12 dB SNR and by 1.5\% at -6 dB SNR, suggesting that DNN-HA based hearing enhancement was possible for adverse listening conditions even in this HI case ($>$50\% loss of ANFs). 

\subsection{Applying the framework to different hearing-loss profiles} \label{sec:simpl}

\begin{figure*}[tb!]
\begin{center}
\includegraphics[width=0.95\textwidth,height=\textheight,keepaspectratio]{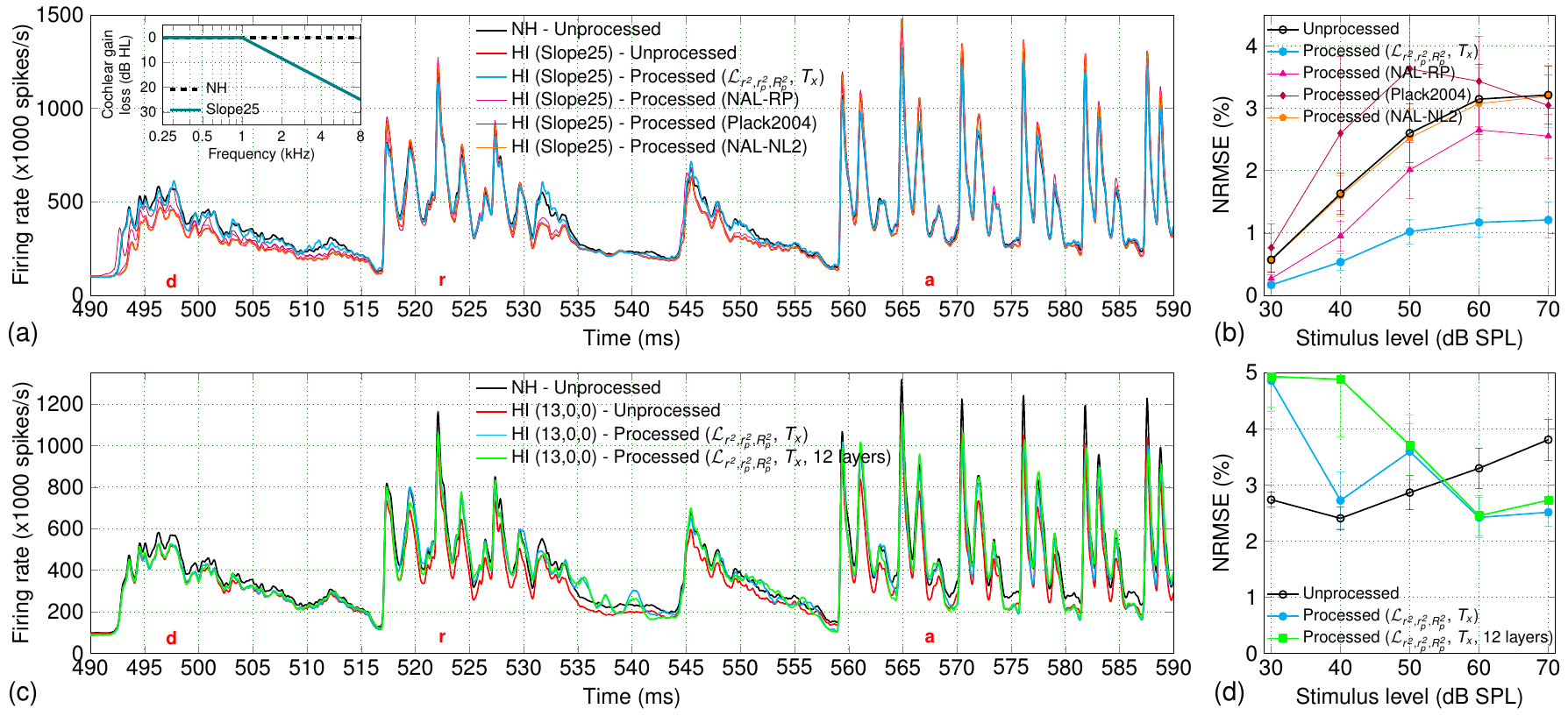}
\end{center}
\vspace{-10pt}
\caption[Restoration of AN population responses for two milder HI cases.]{Restoration of AN population responses for two milder HI cases. Loss function $\La_{r^2,r^2_p,R^2_p}$, $T_{x}$ was used to train two models, to restore simulated AN responses for two milder HI profiles that included only OHC loss (Slope25) and only CS (13,0,0), respectively. Panels (a) and (c) show the simulated enhancement on the AN population responses for an example sentence of the Flemish Matrix. Panels (b) and (d) compare the average NRMSEs between the NH and the respective HI population responses, computed from the 260 sentences of the Flemish Matrix before and after processing. In the case of CS, the trained models were also able to increase the EFR\textsubscript{sum} metric (Eq.~\ref{eq:EFRs}) to $\sim$7.7 nV after processing (6.95 nV before processing).}
\label{fig:ANs_simpl}
\vspace{-10pt}
\end{figure*}

To further explore the restoration capabilities of the best-performing loss function ($\La_{r^2,r^2_p,R^2_p}$, $T_{x}$), we applied the same training procedure to two less severe HI profiles: The first consisted of a milder sloping high-frequency OHC loss (starting from 1 kHz with 25 dB HL at 8 kHz; inset in Fig.~\ref{fig:ANs_simpl}(a)) without CS, abbreviated as \emph{Slope25}. The second included only mild ANF damage ($H\textsubscript{HI}$ = 13, $M\textsubscript{HI}$ = 0, $L\textsubscript{HI}$ = 0) without OHC loss, abbreviated as \emph{13,0,0}. Figure~\ref{fig:ANs_simpl} visualises the restoration capabilities of the trained DNN-HA models on simulated AN population responses, with the selected loss function achieving an accurate compensation for OHC loss (Fig.~\ref{fig:ANs_simpl}(a)).
Three amplification strategies from the openMHA toolbox \cite{kayser2022open} are plotted alongside the DNN-HA results for comparison: The linear NAL-RP prescription strategy \cite{byrne1990hearing}, and the Plack2004 \cite{ewert2011model,plack2004inferred} and NAL-NL2 \cite{keidser2011nal} compressive strategies.
Our trained model outperformed all amplification strategies for clean speech, resulting in average NRMSEs of $<$1\% (Fig.~\ref{fig:ANs_simpl}(b)) without showing over-amplification of the simulated AN responses beyond the NH peaks, as was the case for the Plack2004 strategy (Fig.~\ref{fig:ANs_simpl}(a)). On the other hand, the NAL-NL2 strategy provided only modest amplification for the selected HI profile (Supplementary Fig.~\ref{fig:gaintables}) which could not achieve significant NRMSE improvements after processing, and was thus excluded from further analysis.
The enhancement differences of our trained model and the NAL-RP and Plack2004 amplification strategies are also visualised across time and frequency in Supplementary Fig.~\ref{fig:AN_slope25}. Our DNN-HA model matched the patterns in the tails of the two consonants more accurately than the two standard HA strategies, and this shows promise for unconstrained, end-to-end HA processing algorithms that do not rely on an a priori frequency analysis of the signal.

In case of CS, only a partial restoration of the AN population responses was observed after processing (Fig.~\ref{fig:ANs_simpl}(c)). 
The trained model ($\La_{r^2,r^2_p,R^2_p}$, $T_{x}$) improved the NRMSE for stimuli presented at 60 and 70 dB SPL by $\sim$1\%, but resulted in larger errors for lower speech levels (30-50 dB SPL). 
Although we found that 16 layers achieved the best improvement for the Slope35-7,0,0 HI profile (see Appendix B), a smaller DNN-HA architecture with fewer trainable parameters could show an improved benefit for this milder HI profile (loss of LSR and MSR ANFs only) which might require simpler sound processing. To test this, we applied the same training procedure using an architecture of 12 layers (6 in the encoder and 6 in the decoder) and saw that the smaller architecture ($\La_{r^2,r^2_p,R^2_p}$, $T_{x}$, 12 layers) achieved similar enhancement of the AN population responses (Fig.~\ref{fig:ANs_simpl}(c)) but resulted in higher NRMSEs compared to the original $\La_{r^2,r^2_p,R^2_p}$, $T_{x}$ model (Fig.~\ref{fig:ANs_simpl}(d)).

The difficulty of restoring CS-compromised peripheries is apparent from these two milder HI cases, since a comparison of the unprocessed NRMSEs in these HI cases (Fig.~\ref{fig:ANs_simpl}(b),(d)) shows much larger differences for the CS profile even at 30 dB SPL ($\sim$3\% compared to $\sim$0.5\% for the OHC-loss case).
This suggests that fully restoring CS using the proposed method is much more difficult than the restoration of OHC loss, and especially for severe ANF losses (e.g. $>$50\% loss of ANFs as in the previous section). 
Even though our CS strategies can optimally drive the remaining ANFs after processing, they could not fully compensate for a sheer loss of the ANF population.
This points to a general limitation of acoustic-based treatments for damaged AN structures. 

\begin{figure*}[tb!]
\begin{center}
\includegraphics[width=0.95\textwidth,height=\textheight,keepaspectratio]{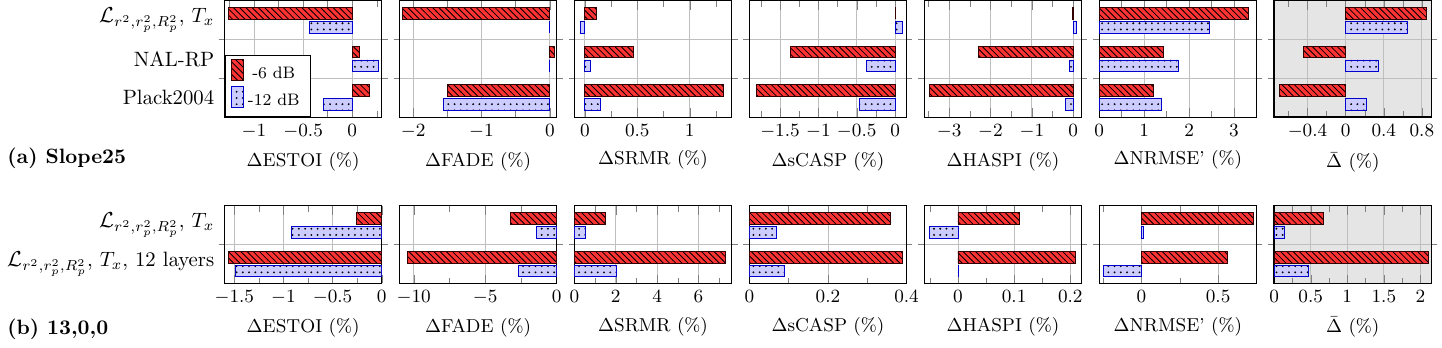}
\end{center}
\vspace{-10pt}
\caption[Assessing speech-in-noise benefit for two milder HI cases.]{Assessing speech-in-noise benefit for two milder HI cases. The average percent improvement on the different metrics is given for the trained models of the two HI profiles: (a) Slope25 and (b) 13,0,0. The percent improvement of the normalised metrics and of the $\bar{\Delta}$ metric (Eq.~\ref{eq:Dm}) was computed for the 70-dB-SPL Flemish Matrix sentences at -6 and -12 dB SNR (cf. Fig.~\ref{fig:optim}). For the HASPI metric, HL of 8.33, 16.66 and 21.54 dB was included at 2, 4 and 6 kHz for the OHC-loss profile (Slope25), while no HL was included for CS evaluation (13,0,0).}
\label{fig:simpl}
\vspace{-10pt}
\end{figure*}

Finally, Fig.~\ref{fig:simpl} shows how well the trained models performed on speech in noise.
In case of OHC loss (Fig.~\ref{fig:simpl}(a)), the two standard amplification strategies \cite{byrne1990hearing, ewert2011model} showed benefits in the noisiest scenario (-12 dB SNR), but the DNN-HA model ($\La_{r^2,r^2_p,R^2_p}$, $T_{x}$) still outperformed both strategies with a $\bar{\Delta}$ benefit of $\sim$0.65\% at -12 dB SNR and of $\sim$0.85\% at -6 dB SNR. 
The DNN-HA model decreased the NRMSEs at $\sim$2.4\% after processing (Supplementary Table~\ref{tab:simpl_all}) but did not reach the NRMSE scores for processed speech in quiet (Fig.~\ref{fig:ANs_simpl}(b); $\sim$1.1\% at 70 dB SPL), suggesting that the restoration of speech in noise was more difficult than for clean speech.
A similar trend was observed for the 13,0,0 CS model ($\La_{r^2,r^2_p,R^2_p}$, $T_{x}$), with the NRMSE benefit gradually diminishing from $\sim$1.3\% in quiet (Fig.~\ref{fig:ANs_simpl}(d)) to 0.7\% at -6 dB SNR and to 0.1\% at -12 dB SNR (Fig.~\ref{fig:simpl}(b)). 
However, the smaller architecture ($\La_{r^2,r^2_p,R^2_p}$, $T_{x}$, 12 layers) showed a significant benefit on the considered speech-intelligibility metrics, improving the SRMR metric by $\sim$7.3\% for speech at -6 dB SNR (Fig.~\ref{fig:simpl}(b)).
Lastly, when it comes to predicted speech quality in noise, we report that both Slope25 and 13,0,0 CS trained models increased the estimated PESQ scores after processing (Supplementary Table~\ref{tab:simpl_all}).

\section{Discussion}

We presented a fully differentiable framework that can train DNN-HA models via backpropagation to minimise the difference between simulated AN responses of NH and HI peripheries, offering a novel type of end-to-end audio processing for hearing aids. The encoder-decoder DNN-HA architecture was fast to execute, with the trained models requiring $\sim$56 ms for the processing of the Flemish Matrix sentences (average length of 2.7 s) on a CPU (AMD EPYC 7413 24-Core) and $\sim$17 ms on a GPU (NVIDIA A30) on average.
Both the encoder and decoder parts of the chosen architecture consisted of 8 convolutional layers with strides of 2, thus the trained DNN-HA models can process inputs with a minimum size of $2^8$ = 256 samples (12.8 ms). At the same time, even though frames of 2048 samples (102.4 ms) were used for training, the input stimuli were not sliced for the evaluation of the models. This demonstrates that the trained DNN-HA models can effectively process inputs of any size due to the nature of their learned operations (convolutions across time), with only small degradations expected in performance for inputs of smaller size than the one used for training \cite{drakopoulos2023icassp}. Architectures with less (strided) layers in the encoder will further decrease the minimum frame size and associated latency, allowing for direct embedding of such models in HA devices that require latencies below 10 ms \cite{balling2020reducing}.
Although we found that an architecture with 16 CNN layers was the best choice for severe HI (Appendix B), smaller encoder-decoder architectures (e.g. 5 or 6 layers in the encoder) achieved similar restoration outcomes and could benefit training in cases with less ANF or OHC damage, as shown in Section~\ref{sec:simpl}.
At the same time, the architectures that we used in this work comprised non-causal convolutional layers, thus the future use of causal convolutions could further decrease the execution times.

\subsubsection*{Simulated restoration} Our framework was first developed and applied for a mixed OHC-loss and CS pathology (Slope35-7,0,0), and then to two simpler HI profiles that included only OHC loss (Slope25) and only CS (13,0,0). The trained DNN-HA models enhanced the simulated AN responses in all cases, but were not able to fully restore them to the responses of the NH periphery for the HI profiles that included CS.
This observation is reflected by the small RMSE reduction that the Slope35-7,0,0 and 13,0,0 models achieved after processing (NRMSEs of $\sim$10\% in Table~\ref{tab:clean} and $\sim$3\% in Fig.~\ref{fig:ANs_simpl}d for processed speech in quiet at 30-70 dB SPL). 
Even though the trained DNN-HA models did not fully restore the simulated responses for profiles that additionally included CS, they were able to enhance the simulated EFR magnitudes (see Section \ref{sec:sim_clean}) by increasing them to $\sim$7.7 nV after processing in both HI cases (NH reference of 8.39 nV). This suggests that the DNN-HA models might still be able to restore specific functional aspects that are degraded by CS (e.g. temporal-envelope coding \cite{keshishzadeh2020derived, vasilkov2021enhancing, drakopoulos2022model}).
At the same time, there may be additional benefits for considering CS in HA processing, as HI listeners with similar audiograms can have very different outcomes on aided speech intelligibility \cite{lopez2017predictors}.

Similar improvements were found when adding noise to the Flemish Matrix sentences, with the Slope35-7,0,0 and 13,0,0 models showing small NRMSE reductions at negative SNRs (Figs.~\ref{fig:optim} and ~\ref{fig:simpl}(b); up to $\sim$1.5\% NRMSE improvement after processing). On the other hand, the Slope25 model was able to improve the NRMSE to $\sim$2.4\% of the NH responses in both SNR conditions (-12 and -6 dB) after processing (Supplementary Table~\ref{tab:simpl_all}).
Considering all HI profiles, it is important to note that the NRMSE scores of the unprocessed sentences significantly increased after adding noise (e.g. 26.3\% at -12 dB and 24.2\% at -6 dB SNR for the Slope35-70,0,0 profile, compared to 13.8\% for clean speech). This is because the analytical model \cite{verhulst_hearres2018} requires more than one (e.g. 20) realisations of noisy stimuli to get a stimulus-driven response. Due to the high computational cost, we used only one realisation throughout our evaluation, since we focussed on a comparative analysis of the simulated outcomes (before and after processing). Thus, lower NRMSE scores (and greater benefits) are expected in low SNR conditions when using more noise realisations.

Additionally, although a larger training dataset could improve the restoration accuracy of the DNN-HA models, our previous studies demonstrated that the TIMIT speech corpus is sufficient to train generalisable models that predict well the non-linear responses to diverse stimuli which were not part of training (e.g. pure and modulated tonal stimuli, clicks or music of different levels and frequencies \cite{baby2021convolutional,drakopoulos2022model}). Thus, we do not expect a significant divergence from our presented results when another dataset is used for training or evaluation. However, the inclusion of datasets with a sufficiently high sampling frequency and broader frequency content (e.g. noise or music) in the training would ensure good model accuracy up to frequencies of 8 kHz.

\subsubsection*{Objective performance} The objective perceptual metrics that we used for the evaluation of speech in noise showed only small speech-intelligibility benefits in most cases, especially for the more complex HI profile (Slope35-7,0,0). This was expected, since no metric took into account the functional effects of CS, and only the HASPI model could include individual OHC-loss profiles in its predictions. At the same time, most metrics used the clean speech signal as the (ideal) reference (Table~\ref{tab:metrics}), even though the evaluation of hearing-restoration strategies should be done by comparing the performance of a HI listener against the reference performance of NH listeners. 
Thus, even though the trained DNN-HA models failed to improve the scores of most considered metrics (reflecting the relative performance of NH listeners in noise), they could still improve speech intelligibility when applied to listeners with OHC loss and/or CS.
This shows the necessity of future subjective evaluation where we can specifically assess the benefit of individualised DNN-HA models that compensate for individual HI profiles.

From the eight considered speech-intelligibility metrics, the SRMR, sCASP and HASPI models proved the most robust metrics for predicting intelligibility after processing, showing small but consistent improvements for the DNN-HA models.
Although the use of objective metrics for non-linear processing strategies that were not explicitly part of the metric development might provide inaccurate predictions, we chose to use these metrics as a first indicator of the perceptual benefits that our DNN-HA strategies can achieve. At the same time, we chose to perform our evaluation based on the Flemish Matrix material \cite{luts2014development} and used realistic SNR conditions (-12, -6 and 0 dB). Thus, a future subjective evaluation of the developed DNN-HA strategies on the Flemish Matrix test will enable a direct comparison between measured speech intelligibility and the objective intelligibility scores that were shown here. By comparing measured and simulated intelligibility results across the presented metrics, we will be better able to assess the reliability of each objective metric for predicting human performance when developing the newest generation of DNN-based non-linear processing strategies.
Given the limited usability of the existing objective metrics for our use cases, the best way to extend the simulated outcomes to perceptual benefits for the trained DNN-HA models is by connecting a speech-intelligibility backend to our periphery model that can accurately predict psychoacoustic performance of NH and HI listeners \cite{schadler2016simulation, haro2020deep}. When sufficient subject data becomes available, a computational backend (e.g. DNN-based) could be developed to predict speech intelligibility from simulated AN responses \cite{haro2020deep}. 

\subsubsection*{Training procedure} For the mixed-pathology HI profile (Slope35-7,0,0), the loss function that gave the best results was the $\La_{r^2,r^2_p,R^2_p}, T_{x}$ (Fig.~\ref{fig:optim}). This loss function used squared time-domain representations and power spectrograms ($\La_{r^2,r^2_p,R^2_p}$) to minimise the AN summed and AN population responses during training (Table~\ref{tab:models}), both in the time and the frequency domain. We found that including STFT-based spectrogram representations in the loss functions benefitted the training (Appendix B) by allowing the DNN-HA models to introduce phase differences in the responses after processing. At the same time, Fig.~\ref{fig:ANsummed} (and Supplementary Fig.~\ref{fig:AN}) demonstrated that models trained with non-squared losses focussed more on the enhancement of the less-pronounced regions of the responses, while models trained with squared losses yielded a better enhancement of the temporal peaks of the responses. Thus, the former models are expected to yield better enhancement of unvoiced and stop constants, and the latter better enhancement of vowels or voiced consonants of speech \cite{drakopoulos2022model}.

At the same time, the best speech-in-noise performance was achieved by adding a stimulus threshold $T_x$ in loss function $\La_{r^2,r^2_p,R^2_p}$ (Appendix B), to omit the silent parts of the inputs during training and focus on the optimisation of the relevant regions of the responses. Although this thresholding targeted response parts that corresponded to the stimulus and benefitted training, it could possibly affect the compensation of low-level stimuli in noisy datasets if not chosen correctly. Thus, more precise techniques, such as a differentiable voice-activity detector, could be used instead of an RMS threshold to further improve the trained aspects and provide a more generalisable restoration.

\subsubsection*{Limitations and outlook} In terms of the DNN-HA performance on simulated AN responses, an almost complete restoration of the AN population responses was achieved for OHC loss alone (Fig.~\ref{fig:ANs_simpl}(a)) with an RMSE reduction to $\sim$1\% of the maximum AN firing rate after processing (Fig.~\ref{fig:ANs_simpl}(b)).
This is an important outcome because it shows that our DNN-based method performs on par with state-of-the-art HA algorithms that focus on compensating for OHC damage. 
The HASPI model was the only metric that included the individual HL to estimate speech intelligibility, and predicted better intelligibility improvements in noise for our DNN-HA model compared to two conventional amplification strategies from the openMHA toolbox \cite{kayser2022open} (Fig.~\ref{fig:simpl}(a)).
Therefore, our framework holds promise for a novel end-to-end and precise treatment of OHC loss, whereas it is not yet fully clear to which degree we can compensate for CS loss.
Regrowth strategies might be necessary together with acoustic compensation to effectively restore the affected AN responses.

Specifically, we found that even mild cases of CS resulted in AN responses that differed significantly from the NH responses even at low stimulus levels (Fig.~\ref{fig:ANs_simpl}(d)), suggesting that it might not be possible to fully restore peripheral coding after a substantial decrease in ANF population, even though our method is successful in more efficiently driving the remaining ANFs. 
This goes in line with the results of our previous work \cite{drakopoulos2022differentiable}, which showed that more precise improvements were achieved for a cochlear gain loss of 25 dB than for a severe case of CS. 
A possible explanation for this could also be the way that CS is introduced in a HI periphery. To simulate CS, the number of ANFs after the simulated outputs is reduced (Fig.~\ref{fig:fw}(b)), resulting in AN responses that are summed across fewer fibres. 
This scaling of ANFs could reduce the degrees of freedom in the earlier parts of the periphery (cochlea and IHC), creating an imbalance between the NH and HI pathways and making the training convergence more difficult.
At the same time, although the simulation of CS was limited to linear adjustments of the number of ANFs in this work, frequency-dependent ANF loss (e.g. high-frequency sloping CS) or different types of neural combination of the ANF population (e.g. modulation filters that approximate the cochlear nucleus and inferior colliculus auditory-processing stages \cite{dau2003importance, bourien2014contribution, verhulst_hearres2018}) could simplify the problem and result in better restoration.
The future inclusion of detailed models of auditory brainstem processing in our framework could also account for central auditory-processing changes after the AN that can follow SNHL (e.g. compensatory mechanisms such as central gain).

The differentiable framework we proposed here can directly be used to design preset hearing-enhancement strategies, or OHC-loss compensation strategies based on the measured audiogram of a HI listener. Since the Slope35-7,0,0 HI combination was chosen to reflect the average predicted SNHL profile among older NH people \cite{keshishzadeh2021towards}, the trained DNN-HA models could be used to improve peripheral sound encoding in older listeners who experience hearing difficulties but are clinically considered as normal hearing and are thus currently left untreated.
Predesigned DNN-HA strategies can also be evaluated with HI listeners and compared to conventional HA strategies (e.g. NAL-NL2 \cite{keidser2011nal}) on the basis of standardised listening tests that include aspects such as listener preference, speech intelligibility and quality.
Following a diagnostic session where the SNHL profile of a listener is determined in detail, the HI periphery in our framework can also be adjusted based on the individual degree of OHC loss and CS to design a hearing-restoration strategy tuned to the individual listener \cite{keshishzadeh2021towards}. Then, the developed DNN-HA model can be evaluated experimentally with the same listener, to evaluate the benefit that the processing has on physiological and behavioural markers (e.g. EFRs, AM detection and word recognition in noise \cite{drakopoulos2022model}).

Although the presented framework can only be used to design individualised or off-the-shelf HA strategies in its current form, further adaptations to the framework can provide more generalisable solutions that can be used by multiple individuals (e.g. parameterised models \cite{bystedparameter} or adjustable HA processing based on a provided ``hearing-loss'' input). Along those lines, we recently proposed a modified training procedure that considers different OHC-loss profiles in the framework to train a single DNN-HA model architecture \cite{drakopoulos2023icassp}. The resulting DNN-HA model can thus provide individualised HA processing based on an audiogram input after training, without having to retrain the model parameters for an individual listener\footnote{The trained DNN-HA model is available via \protect\url{https://doi.org/10.5281/zenodo.7717218} or \protect\url{https://github.com/HearingTechnology/DNN-HA}.}.

Finally, we applied several constraints in the loss functions with the aim of providing a more principled enhancement of the simulated responses, but we refrained from applying any constraints to the sound processing that the DNN-HA models provide. Even if the DNN-HA models improve intelligibility for impaired listeners after training, such non-linear processing can often lead to perceivable sound distortions or poor quality \cite{moore2008choice}. However, the estimated PESQ scores for the different strategies improved in most cases after processing (Supplementary Tables~\ref{tab:optim_all},\ref{tab:simpl_all}), which suggests that no significant distortions were introduced by our processing strategies. Although we refrained from including an in-depth analysis of the non-linear sound processing that the trained DNN-HA models applied in this paper, our preliminary work has demonstrated how such an analysis can be performed for DNN-based speech processing \cite{wouters2022hearing}. Future work is planned that will thoroughly assess the characteristics of the non-linear sound processing of the trained DNN-HA models, and the effect that this processing has on measured sound quality and speech intelligibility in humans.

\section{Conclusion}

We presented how a differentiable neural-network framework can be used to design individualised hearing-restoration strategies. The framework was based on CoNNear \cite{baby2021convolutional,Drakopoulos2021}, a CNN-based model of human auditory processing, and was used to train DNN-based HA models that can minimise the difference between simulated AN responses of a NH and a HI periphery. Different loss functions were evaluated and compared in terms of simulated restoration and objective speech-intelligibility benefit in noise.
Overall, the trained DNN-HA models improved the simulated outcomes for speech in quiet, and showed benefits on the RMSE between the NH and HI responses and on different speech-intelligibility metrics for the Flemish Matrix \cite{luts2014development} sentences at negative SNRs. 
The framework was more performant for OHC loss compensation than for CS, accurately restoring the simulated AN population responses and achieving a significant RMSE reduction. A complete restoration of the simulated responses was not possible in the cases of HI peripheries that included CS, and this suggests that acoustic-based treatments might not be able to fully compensate for damage of AN structures. More constrained training approaches or perceptually relevant loss functions might help achieve truly individualised hearing restoration in these cases. 
We believe that our framework can pave the way for precision treatment of SNHL and can be used to develop the next generation of hearing aids.

\bibliography{refs}
\bibliographystyle{IEEEtran}


\section{Biography Section}

\begin{IEEEbiography}[{\includegraphics[width=1in,height=1.25in,clip,keepaspectratio]{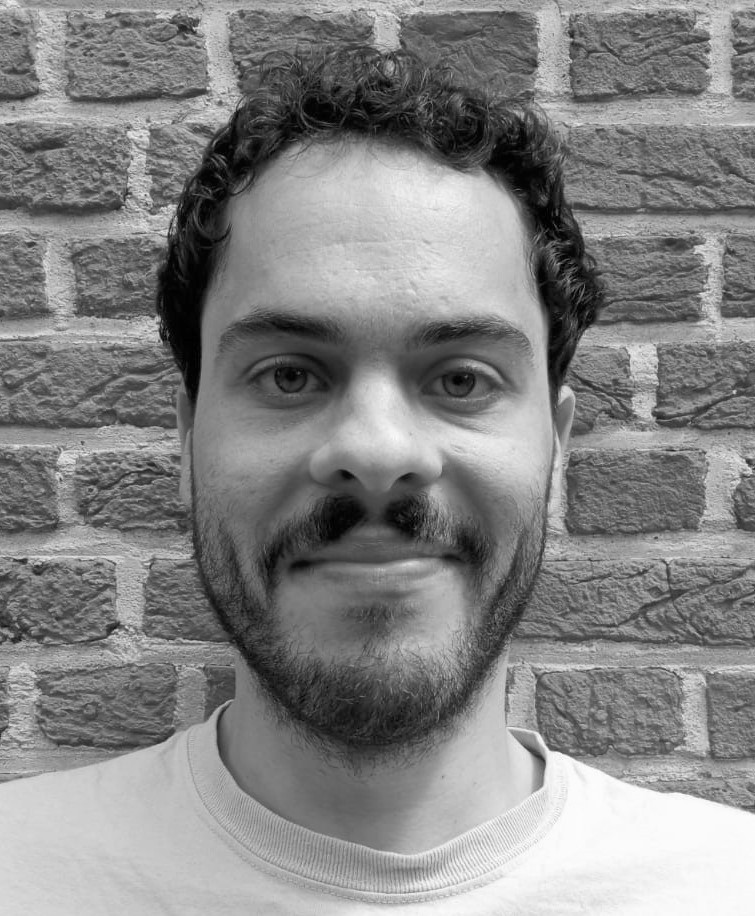}}]{Fotios Drakopoulos}
(Member, IEEE) received the Ph.D. degree in electrical engineering from Ghent University, Ghent, Belgium, as part of the Hearing Technology Lab led by Prof. Sarah Verhulst, in November 2022. He is currently a Postdoctoral Researcher with the Ear Institute of University College London, London, U.K. He received the M.Sc. degree from the electrical and computer engineering department at University of Patras, Greece, in 2018, under the supervision of Prof. John Mourjopoulos. He was an Intern in Phonak (Sonova AG) with Dr. Eleftheria Georganti. His main research interests include audio signal processing, auditory modeling, machine hearing, hearing devices, and psychoacoustics. He was a Reviewer of \textit{Nature Communications Biology}, \textit{Journal of the Acoustical Society of America} (JASA), and \textit{Trends in Hearing}.
\end{IEEEbiography}

\begin{IEEEbiography}[{\includegraphics[width=1in,height=1.25in,clip,keepaspectratio]{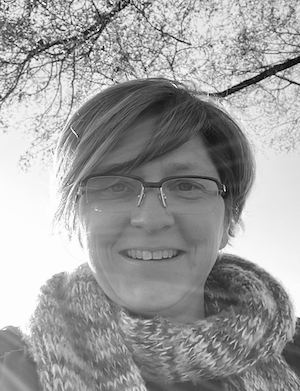}}]{Sarah Verhulst}
(Member, IEEE) received the Ph.D. degree from the Technical University of Denmark in 2010. She is Full Professor with Ghent University, Belgium, and has been the Head of the Hearing Technology Lab since 2016. She has a background in electrical and acoustical engineering. After Postdoctoral training with Boston University and Harvard, Boston, MA, USA, she became an assistant Professor with Oldenburg University, Germany. She holds an ERC-StG, ERC-PoC and EIC-transition grant (end dates 2021, 2022, 2025) and was PI in several large-scale research networks over the past years (DFG Priority program 1608 and ERA-NET Neuron). Her team with Ghent University takes a multi-disciplinary experimental/theoretical approach to develop hearing technologies, with focus on auditory modelling, signal processing for EEG-based hearing diagnostics, and machine-learning methods for hearing-aid signal processing. She was the recipient of the 2016 Niedersachsen Wissenschaftspreis (Kat.II), the 2018 Young Investigator Spotlight award from the auditory neuroscience community (APAN) and 2022 Laureate of the KVAB (Technical Class). She is member of the Belgian Young Academy, and an Associate editor for the \textit{Journal of the Acoustical Society of America} (JASA).
\end{IEEEbiography}


\vfill

\clearpage 
\newpage

\renewcommand{\figurename}{Supplementary Fig.}
\renewcommand{\tablename}{Supplementary Table}
\setcounter{figure}{0}
\setcounter{table}{0}

    
    
    
    


\begin{figure*}[htb]
\begin{center}
\includegraphics[width=\textwidth,height=\textheight,keepaspectratio]{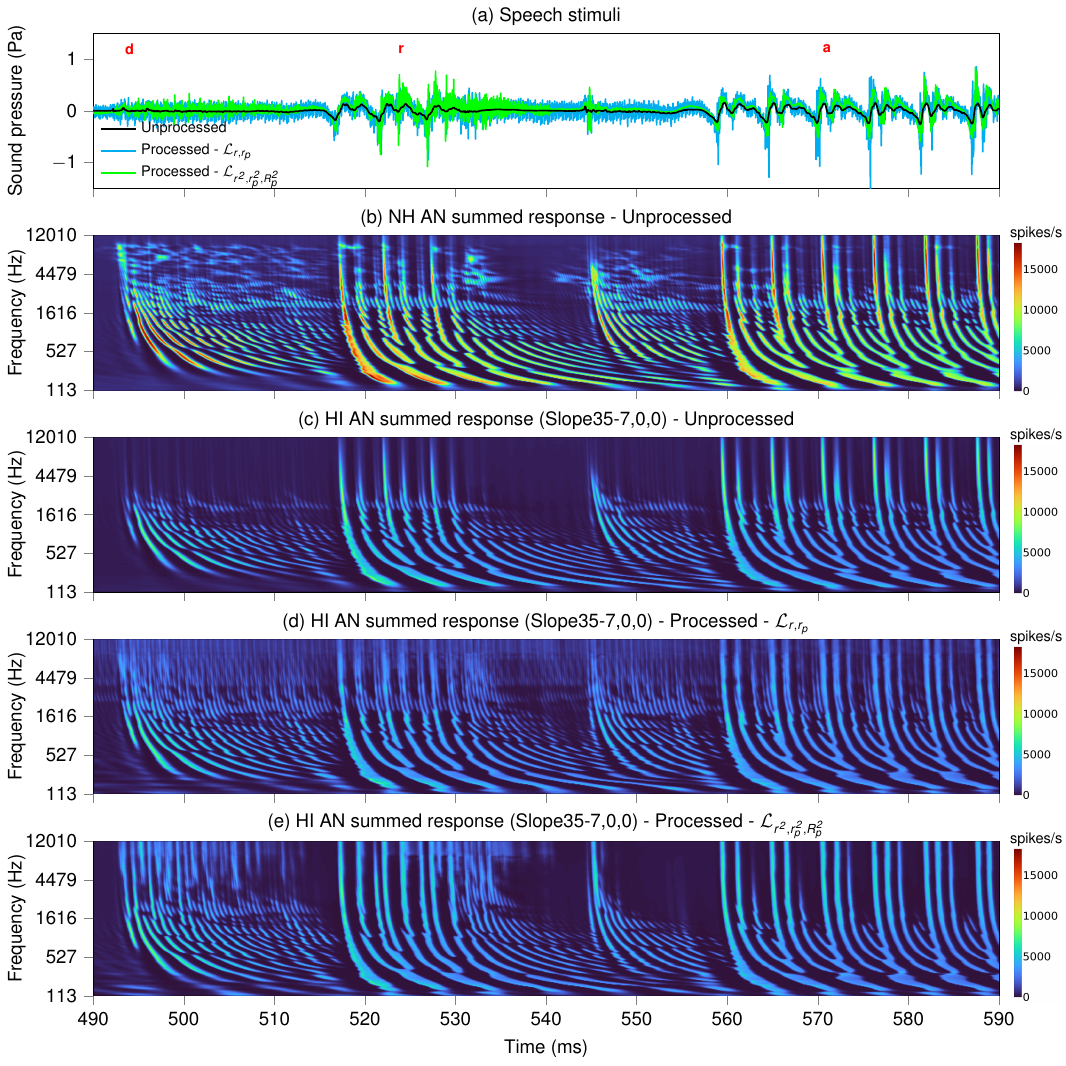}
\end{center}
\vspace{-10pt}
\caption[Simulated AN responses for two trained DNN-HA models (Slope35-7,0,0).]{Simulated AN responses for two trained DNN-HA models (Slope35-7,0,0). (a) A speech segment is shown before and after processing with models $\La_{r,r_p}$ and $\La_{r^2,r^2_p,R^2_p}$, containing two consonants (\textbackslash d\textbackslash , \textbackslash r\textbackslash) and a vowel (\textbackslash a\textbackslash ) from the Flemish word `draagt'. The unprocessed stimulus was used to simulate the AN summed response of the NH (b) and the HI (c) periphery. The processed stimuli of the two DNN-HA models were given to the HI periphery, resulting in different enhancement of the AN responses (panels (d),(e)). Model $\La_{r^2,r^2_p,R^2_p}$ (e) enhanced the most pronounced parts of the AN responses while leaving the unexcited regions intact, yielding a stronger onset afterwards (e.g. 520-530~ms). The generation of stronger onset responses after increased silence periods is a typical property of the AN fibres \cite{vasilkov2021enhancing, drakopoulos2022model}, and it is interesting to see that the DNN-HA processing managed to exploit this feature. On the other hand, model $\La_{r,r_p}$ relied more on the stimulation of the previously unexcited high frequencies (channels 0-70, frequencies above $\sim$2~kHz) to yield an enhanced AN population response (d).}
\label{fig:AN}
\end{figure*}

{\appendices
\section{Objective metrics}

To evaluate the simulated benefits of the trained DNN-HA models for speech in noise, we used nine well-established objective metrics (Table~\ref{tab:metrics}).
Most of these metrics predict speech performance by comparing a distorted signal (e.g. noisy, reverberated, processed) to a reference signal (usually the clean signal before noise or processing).
Therefore, we performed the objective evaluation of the PESQ \cite{rix2001perceptual}, STOI \cite{taal2011algorithm, jensen2016algorithm}, SIIB \cite{van2017instrumental}, sEPSM\textsuperscript{corr} \cite{relano2016predicting}, sCASP \cite{relano2019speech} and HASPI \cite{kates2021hearing} metrics by comparing the noisy sentences of the Flemish Matrix (before or after processing) to the unprocessed clean (noise-free) sentences. 
For the STOI metric, we evaluated both the regular \cite{taal2011algorithm} and extended \cite{jensen2016algorithm} versions (Fig.~\ref{fig:scores}(b)).
The SIIB metric compared the noisy sentences (before or after processing) to the unprocessed clean sentences, but required a concatenation of all sentences for a reliable comparison. 
The mr-sEPSM model \cite{jorgensen2013multi}, the correlation-based sEPSM\textsuperscript{corr} and the sCASP model predicted speech intelligibility based on the SNR of the noisy speech in the envelope domain (SNR\textsubscript{env}), with the sCASP model additionally using the internal time-frequency representations of an auditory signal-processing frontend \cite{jepsen2008computational} to estimate the average cross-correlation between the noisy (target) and clean (template) sentences. 

The same principle was followed for the mr-sEPSM metric \cite{jorgensen2013multi}, but with the unprocessed noise signal as reference instead (Table~\ref{tab:metrics}). 
The HASPI model was also used to predict speech intelligibility based on a comparison between the noisy and clean sentences, but additionally provided the option of including HL to estimate intelligibility of HI listeners (HI in Fig.~\ref{fig:scores}(h)). In this case, the values of 11.66, 23.33 and 30.16 dB HL were used at frequencies of 2, 4 and 6 kHz for the first HI profile (Slope35-7,0,0), corresponding to the hearing thresholds (audiogram) of the included OHC-loss profile (Slope35 \cite{verhulst_hearres2018}).
The MATLAB implementation of HASPI v2 was provided by the authors. On the other hand, the SRMR metric \cite{falk2010non, santos2014improved} was non-intrusive (blind estimation) and thus predicted speech quality and intelligibility directly from the noisy material, without requiring a reference for comparison.

Lastly, the FADE framework for human psychoacoustics simulation \cite{schadler2016simulation} was used to simulate word-recognition performance. FADE was trained on the Flemish Matrix sentences mixed with noise at SNRs between -24 and 6 dB. After training, the average scores (percentage of correctly identified words) were estimated for the sentences in noise, before and after processing with the DNN-HA models. Three repetitions were used to derive the percent scores of FADE, with the average of the three results reported in this work.
The FADE model showed 15.44\% word recognition at -12 dB SNR and 91.72\% word recognition at -6 dB SNR (Fig.~\ref{fig:scores}(i)), corroborating the experimental results of previous studies that used the Flemish Matrix test \cite{maele2021variability, drakopoulos2022model} ($\sim$15\% at -11.5 dB SNR, $\sim$85\% at -5.5 dB SNR for young NH subjects).

\vfill\eject

\section{Additional training constraints}

Based on the loss functions of the eight trained DNN-HA models (Table~\ref{tab:models}), additional training constraints were tested to further improve the minimisation procedure and the trained aspects of each DNN-HA model. The different training setups are explained in the following paragraphs, with the results of the newly trained models juxtaposed with the original models in Fig.~\ref{fig:optim}.

\paragraph{Frequency emphasis}

Since the speech corpus mostly contains energy at the lower frequencies, it was expected that the trained DNN-HA models can compensate for the degraded AN population responses by adding more energy to the less-excited frequency components of the stimulus (i.e. high frequencies). 
This was more apparent with loss functions that only targeted the minimisation of time-domain responses, as was visualised for model $\La_{r,r_p}$ in Fig.~\ref{fig:ANsummed}(a) and Supplementary Fig.~\ref{fig:AN}(d). The trained models added high-frequency energy to the previously unstimulated regions to enhance the AN population responses (Supplementary Fig.~\ref{fig:AN}(d)), but this was avoided with the use of squared time representations in the training (loss functions $\La_{r^2}$-$\La_{r^2,r^2_p,R^2_p}$).

For severe cases of HI with substantial loss at the high frequencies (e.g. Slope35-7,0,0), amplifying these frequency regions could be the optimal way to compensate for the degraded AN population responses. However, we also applied an emphasis function to weigh the AN responses based on the CF of each channel, thereby focussing on the enhancement of the low-frequency channels during training. To this end, we used a sigmoid-like function in loss function $\La_{r}$ and multiplied it by the simulated AN responses across frequency to attenuate the higher frequency channels up to 62\%. 
The trained model ($\La_{r}$ + freq. emphasis) showed improvement over the $\La_{r}$ model for the SRMR and the sCASP metrics (Fig.~\ref{fig:optim}), increasing the average percent improvement ($\bar{\Delta}$) at -12 dB SNR. 
However, the improvement was still not sufficient to outperform models $\La_{r^2}$ and $\La_{r^2,R^2}$.

\paragraph{AN response threshold} Although the loss functions $\La_{r^2}$-$\La_{r^2,r^2_p,R^2_p}$ used squared time representations to focus on the minimisation of the temporal peaks of the responses, another way to achieve this is by applying a temporal threshold to the AN responses. To this end, a static or dynamic threshold can be defined during training so that only the response parts that exceed a certain value are minimised in the loss function. For each frequency channel, we used a moving threshold defined from the RMS of the respective AN response, set to 40\% of the range of the response each time. The moving RMS was computed from the AN responses of each CF by computing the RMS of a moving window of 51 samples ($\sim$2.5 ms). Then, the local extrema of the moving RMS were computed using a moving window of 1001 samples. Based on the estimated RMS range of the responses, the moving threshold $T_{r}$ was defined and only the responses that exceeded the threshold were minimised:
  \begin{equation} \label{eq:Th}
T_{r} = 0.4 \cdot (r_{RMS-max} - r_{RMS-min}) + r_{RMS-min},
  \end{equation}
where $r_{RMS-max}$ and $r_{RMS-min}$ are the moving maximum and minimum of the moving RMS of the AN responses $r$.

The threshold was applied in loss function $\La_{r}$, with the trained model ($\La_{r}$ + $T_{r}$) showing a small benefit only for the noisiest scenario (-12 dB SNR) over $\La_{r}$. Once again, using a moving threshold on the AN responses did not outperform model $\La_{r^2}$ (trained in the same way as model $\La_{r}$ but with squared time-domain responses), supporting our choice for including squared representations in the training.

\paragraph{Complex STFT}
The frequency representations that were used in loss functions $\La_{r,R}$, $\La_{r,r_p,R_p}$, $\La_{r^2,R^2}$ and $\La_{r^2,r^2_p,R^2_p}$ corresponded to the STFT magnitude spectrograms of the AN responses (Eqs.~\ref{eq:lstftch} and \ref{eq:lstftsum}). Although we discarded the phase information of the STFT to allow for phase-insensitive compensation (Section~\ref{sec:loss}), both the real and imaginary parts of the STFT representations can also be used during training. To this end, we chose loss function $\La_{r,R}$ and omitted the absolute value of the STFT in Eq.~\ref{eq:lstftsum}.
The trained model ($\La_{r,R}$, complex STFT) further decreased all of the metrics in both SNR conditions, hence the inclusion of phase-insensitive representations of the AN responses seems to benefit the trained restoration aspects. 

\paragraph{ESTOI optimisation} Apart from the minimisation of the simulated AN responses, training can be focussed to achieve a better perceptual benefit. To this end, we used a differentiable implementation of the ESTOI metric on Tensorflow \cite{kolbaek_loss_2020}. During training, an ESTOI loss function $\ell_{estoi}$ was computed between the unprocessed and processed stimulus and was used together with loss function $\La_{r^2}$, multiplied by 2000 to ensure an approximately equal contribution to the other losses. Although the loss function $\La_{r^2}$ + $2000 \cdot \ell_{estoi}$ was used to maximise the ESTOI metric implementation, the resulting model was not able to increase the average ESTOI scores for the Flemish Matrix sentences and further decreased most of the metrics (Fig.~\ref{fig:optim}). This resulted in an average percent improvement ($\bar{\Delta}$) that was slightly lower than the $\La_{r^2}$ loss function, thus the inclusion of a ESTOI loss function to the training was not able to benefit our objective speech-intelligibility assessment.

\paragraph{DNN-HA architecture} The architecture of the DNN-HA model can also be modified to yield better enhancements. Less deep CNN architectures might still be able to yield the same enhancements and could also avoid overfitting to the training dataset. Thus, we used loss function $\La_{r^2}$ and trained two smaller models with the same architecture but less layers: One with 12 convolutional layers (6 in the encoder and 6 in the decoder) and [16, 32, 32, 64, 64, 128] filters in each layer of the encoder, and one with 10 convolutional layers (5 in the encoder and 5 in the decoder) and [16, 32, 32, 64, 64] filters in the encoder. As before, the number of filters in each encoder layer were mirrored in reverse order in the decoder layers. Fig.~\ref{fig:optim} shows that the two trained models ($\La_{r^2}$, 12 layers and $\La_{r^2}$, 10 layers) had similar performance to the $\La_{r^2}$ model, but decreased the $\bar{\Delta}$ scores in the -6 dB SNR condition by $\sim$0.3\%. Although 6 or 5 layers in the encoder might result in the same restoration capabilities, we chose to keep the original CNN architecture (16 layers) for the DNN-HA models of the Slope35-7,0,0 HI profile.

\paragraph{Frequency resolution} To speed up the training procedure, we chose to use 21 frequency channels out of the 201 that are originally simulated by the CoNNear cochlear model. Since this could have impacted the frequency properties of the achieved processing, we also evaluated the effect of increasing the amount of CFs on the trained processing. To this end, we chose to use 41 channels (1 to 201 with a step size of 5) for the AN responses of loss function $\La_{r^2}$. The trained model ($\La_{r^2}$, 41 CFs) showed only small improvements to the FADE and NRMSE results and decreased the other metrics. Thus, 21 CFs seem to be a good compromise for the frequency resolution of the simulated responses.

\paragraph{Noisy dataset} Although we chose to train all models on clean speech but evaluate them in noise, we also studied the effect of adopting a noisy dataset for training instead. To this end, we trained model $\La_{r^2}$ using the same training dataset but with white noise added at random SNRs from -20 to 20 dB. The resulting model ($\La_{r^2}$, noisy dataset) performed worse on most of the metrics and only showed small improvements on the HASPI and NRMSE metrics. Comparable results were achieved using the noisy dataset even at -12 dB SNR, thus the models seem to apply similar processing to noisy speech even when trained on clean speech. This outcome was not surprising, as the DNN-HA aimed to restore peripheral sound processing and it is per se agnostic to the type of sound it receives.

\paragraph{Stimulus threshold} The training dataset we used consisted of sentences of fixed length ($L\textsubscript{c}$ = 81,920 samples) that were zero-padded to be used as inputs to the CoNNear models. These silent parts of the dataset (and also silence between words) were equally considered during training and might have impacted the trained properties of the DNN-HA models, possibly resulting in models that try to compensate for the response differences even when no stimulation is present (resting firing rate).
To ensure that the trained models mostly focus on the enhancement of non-silent parts of speech, we applied a low-value threshold to the input stimuli during training based on the running RMS value. In each training batch, the RMS was computed from the 81920-sample stimulus using a moving window of 101 samples ($\sim$5 ms), and a threshold $T_{x}$ was defined at 1\% of the maximum RMS value. Then, the regions of the stimulus that were below the threshold were determined, and the corresponding parts of the simulated AN responses were not considered in the loss functions during training. 

The stimulus threshold $T_{x}$ was added to the loss functions $\La_{r^2}$, $\La_{r^2,R^2}$ and $\La_{r^2,r^2_p,R^2_p}$ to train three new models. In the first two cases, the trained models $\La_{r^2}$, $T_{x}$ and $\La_{r^2,R^2}$, $T_{x}$ were not able to show improvement over the original loss functions $\La_{r^2}$ and $\La_{r^2,R^2}$, respectively (Fig.~\ref{fig:optim}). However, the trained model $\La_{r^2,r^2_p,R^2_p}$, $T_{x}$ showed improvement over loss function $\La_{r^2,r^2_p,R^2_p}$ for both SNR conditions (1.1\% at -12 dB and 0.8\% at -6 dB SNR on average). 
}

\begin{figure*}[htb!]
{\begin{center}
\includegraphics[width=\textwidth,height=\textheight,keepaspectratio]{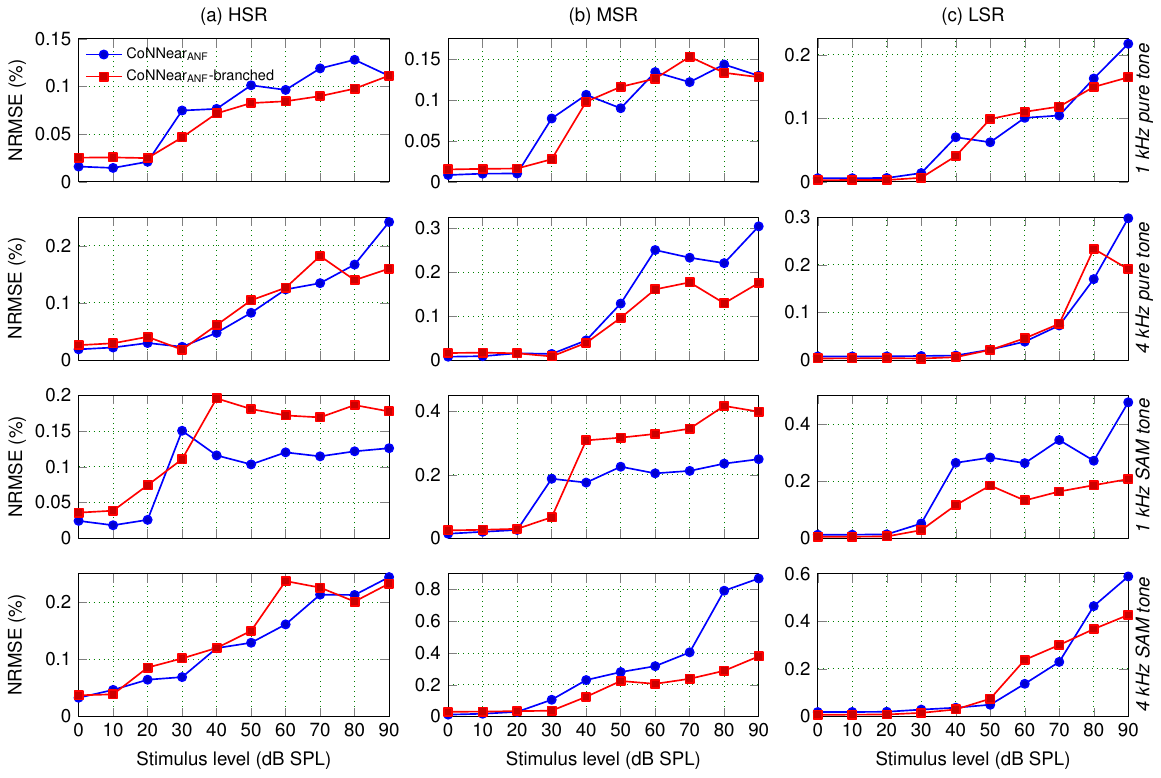}
\end{center}
}
\vspace{-10pt}
\caption[Normalised root-mean-square errors (NRMSEs) between simulated ANF firing rates of the CoNNear ANF model \cite{Drakopoulos2021} and the new branched version.]{Normalised root-mean-square errors (NRMSEs) between simulated ANF firing rates of the CoNNear ANF model \cite{Drakopoulos2021} and the new branched version, shown separately for each fibre type. For each of the tonal stimuli used for the ANF evaluation in \cite{Drakopoulos2021}, the RMSE was computed across time for simulated firing rates of different levels (Supplementary Fig.~3 in \cite{Drakopoulos2021}) and is shown here as a percentage of the maximum value of the respective (reference) firing rate (Fig.~5 in \cite{Drakopoulos2021}). The branched ANF model yielded similar RMSEs for all tonal stimuli, even decreasing the RMSE by $\sim$0.015\% on average.}
\label{fig:rmse_anf}
\vspace{-10pt}
\end{figure*}

\begin{table*}[htb!]
\begin{center}
\caption[CoNNear ANF execution time for a full sentence.]{Comparison of the time required to process the simulated IHC receptor potentials of a full TIMIT sentence using the CoNNear ANF and ANF-branched models on a CPU (Intel Xeon E5-2620 v4 @ 2.10GHz) and a GPU (Nvidia GTX 1080 Ti 12GB). The sentence was zero-padded to account for the context effects of the CoNNear models, and the reported times correspond to the total time needed to transform the $\sim$4.1~s IHC receptor-potential input to ANF firing rates. 201, 21 and 1 channels were selected to demonstrate the effect on the execution time of the ANF models. The single-channel CoNNear\textsubscript{ANF} models were used for all simulations to avoid large memory allocation, and population responses were simulated consecutively (channel by channel). An almost two-fold faster execution was achieved on the CPU and a 1.5-fold speed up on the GPU for the CoNNear\textsubscript{ANF}-branched model.}
\vspace{-5pt}
\centering
\begin{tabular}{c c c c c c c c}
\toprule
\textbf{Model} & \textbf{Window} & \multicolumn{3}{c}{\textbf{CPU (s)}} & \multicolumn{3}{c}{\textbf{GPU (ms)}} \\
\cmidrule(lr){3-5}\cmidrule(lr){6-8}
 & \textbf{(samples)} & \textbf{201-CF} & \textbf{21-CF} & \textbf{1-CF} & \textbf{201-CF} & \textbf{21-CF} & \textbf{1-CF}\\
\midrule
 CoNNear\textsubscript{ANF\textsubscript{H}} & 81,920 & 16.0517 & 1.5811 & 0.0869 & 2746.92 & 259.69 & 14.92 \\
 CoNNear\textsubscript{ANF\textsubscript{M}} & 81,920 & 15.7115 & 1.5865 & 0.0761 & 2613.26 & 242.65 & 11.93 \\
 CoNNear\textsubscript{ANF\textsubscript{L}} & 81,920 & 13.5128 & 1.5733 & 0.0644 & 1895.50 & 181.43 & 8.82 \\
 \hline
 CoNNear\textsubscript{ANF}\text{-branched} & 81,920 & 23.8270 & 2.6655 & 0.1289 & 5504.60 & 544.07 & 23.57 \\

\bottomrule
\end{tabular}
\label{tab:timing}
\end{center}
\end{table*}

\begin{figure*}[bp!]
{\begin{center}
\includegraphics[width=\textwidth,height=\textheight,keepaspectratio]{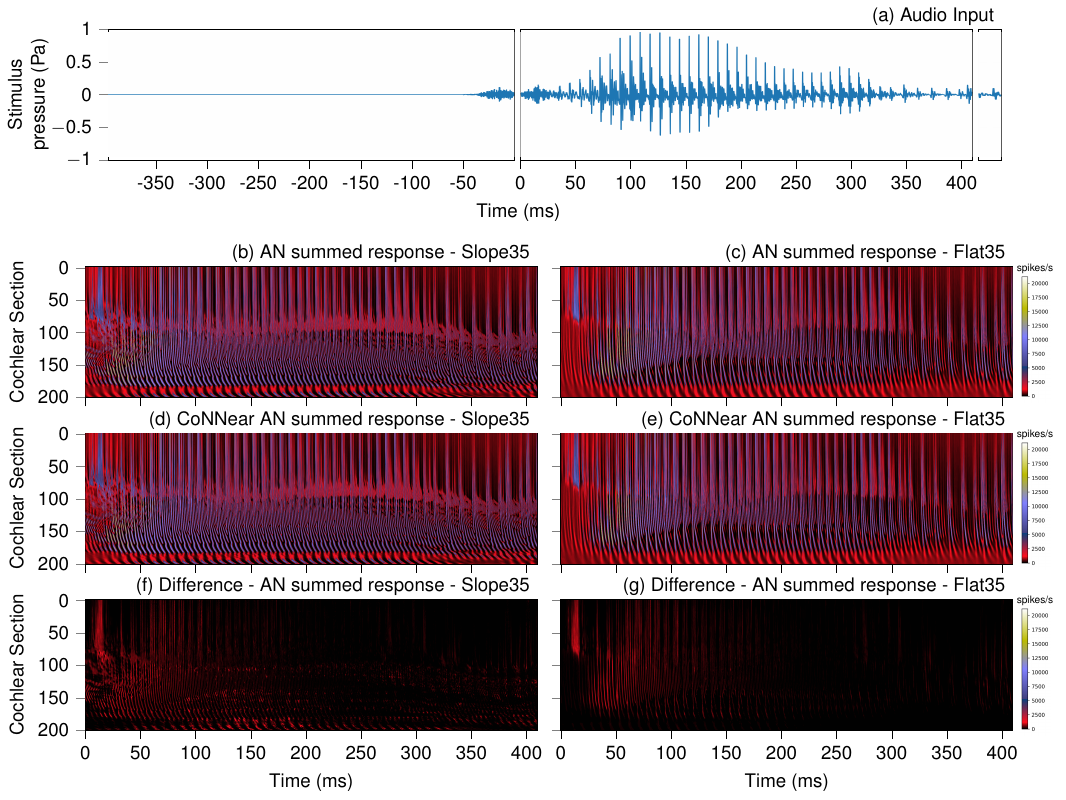}
\end{center}
}
\vspace{-10pt}
\caption[Simulated ANF firing rates for a 8192-sample-long speech stimulus.]{Simulated ANF firing rates for a 8192-sample-long speech stimulus. The stimulus waveform of panel (a) was given as input to the reference auditory periphery model \cite{verhulst_hearres2018} and the CoNNear auditory periphery model (cf. Fig~\ref{fig:fw}(b)) to derive the simulated AN summed responses. Two HI profiles were used: A sloping high-frequency cochlear gain loss of 35 dB at 8 kHz (Slope35), and a flat cochlear gain loss of 35 dB (Flat35). Panels (b) and (c) depict the AN summed responses (in spikes/s) of the reference HI model, while panels (d) and (e) depict the same responses simulated by the HI CoNNear models. Furthermore, panels (f) and (g) show the absolute difference between the simulated outputs of the reference model and the CoNNear model for the two HI profiles. Differences between the reference and CoNNear AN summed responses were mostly found at the stimulus onsets (0-75~ms), with average RMSEs of 1.05\% and 0.79\% for the Slope35 and Flat35 profiles, respectively, compared to the maximum of the AN summed response. }
\label{fig:speech}
\vspace{-10pt}
\end{figure*}

\begin{figure*}[htb]
\begin{center}
\includegraphics[width=\textwidth,height=\textheight,keepaspectratio]{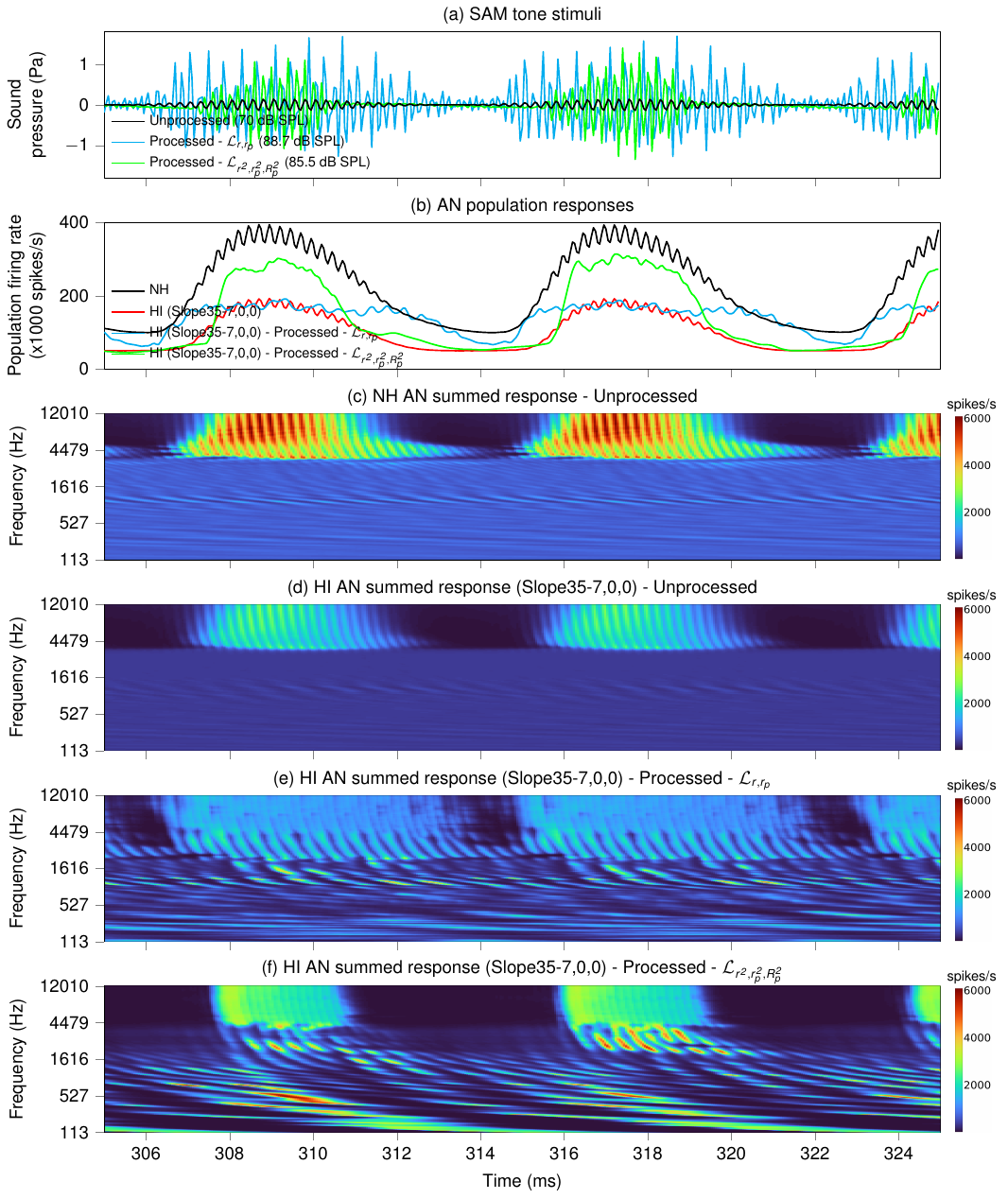}
\end{center}
\vspace{-10pt}
\caption[Simulated AN responses to a SAM tone stimulus (Slope35-7,0,0).]{Simulated AN responses to a SAM tone stimulus (Slope35-7,0,0). (a) The SAM tone stimulus that was used for the simulation of EFRs is shown before and after processing with models $\La_{r,r_p}$ and $\La_{r^2,r^2_p,R^2_p}$. The unprocessed stimulus was used to simulate the AN summed response of the NH (c) and the HI (d) periphery. The processed stimuli of the two DNN-HA models were given to the HI periphery, resulting in different enhancement of the AN responses (panels (e),(f)). For the same stimuli, panel (b) also shows the corresponding AN population responses. Model $\La_{r^2,r^2_p,R^2_p}$ enhanced AN population responses mainly at the stimulus peaks, while model $\La_{r,r_p}$ amplified the signal to yield a more uniform improvement of the firing rate over time.}
\label{fig:AN_SAM} 
\end{figure*}

\begin{figure*}[htb]
\begin{center}
\includegraphics[width=\textwidth,height=\textheight,keepaspectratio]{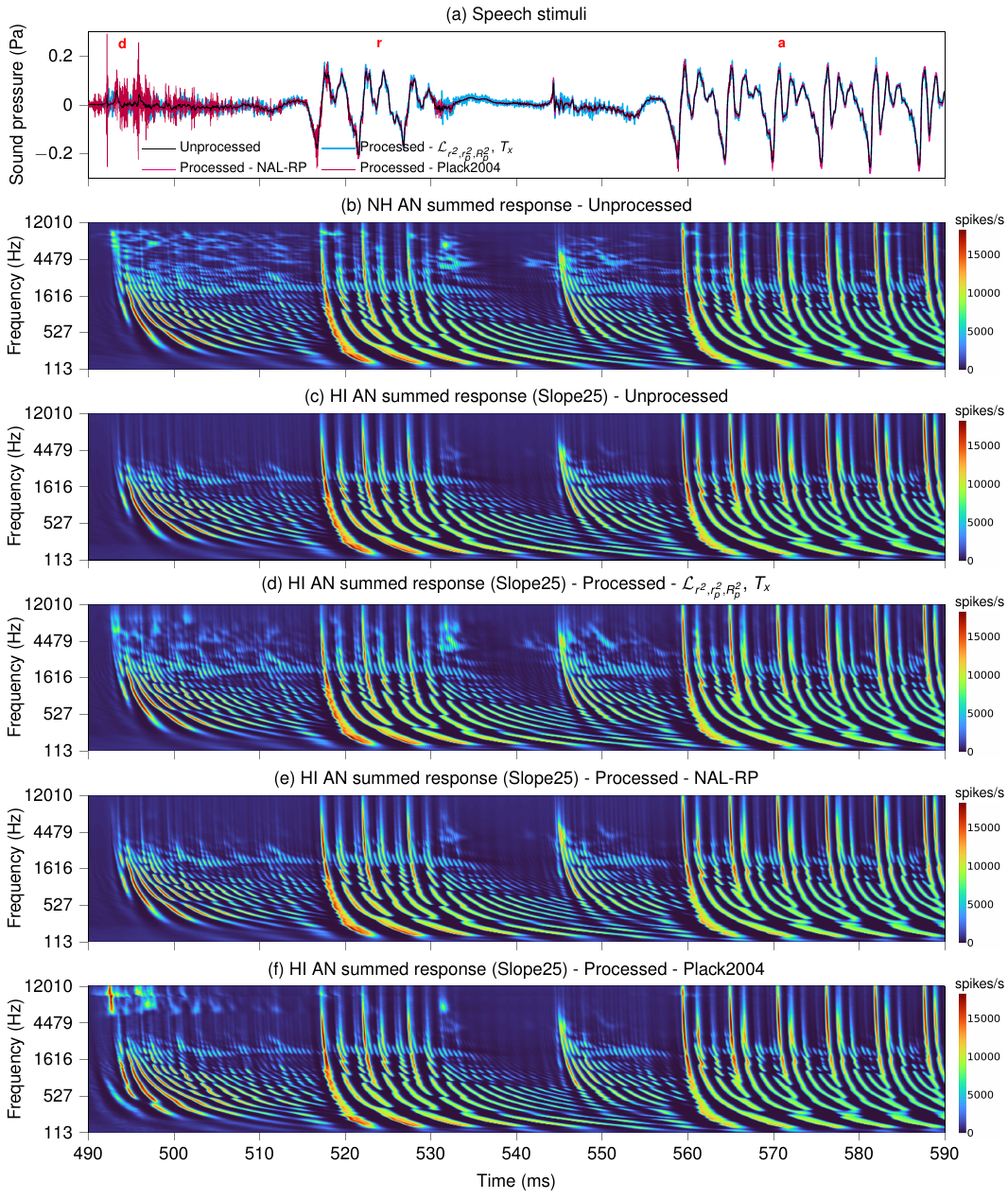}
\end{center}
\vspace{-10pt}
\caption[Restoration of simulated AN responses for OHC loss (Slope25).]{Restoration of simulated AN responses for OHC loss (Slope25). (a) A speech segment is shown before and after processing with model $\La_{r^2,r^2_p,R^2_p}$, $T_{x}$ and two conventional amplification strategies \cite{byrne1990hearing,ewert2011model}. The unprocessed stimulus was used to simulate the AN summed response of the NH (b) and the HI (c) periphery. The processed stimuli of the DNN-HA model and the two HA amplification strategies were given to the HI periphery to compensate for the degraded AN responses (panels (d)-(f), respectively). Although our trained DNN-HA model accurately restored the high-frequency parts of the two consonants (490-560~ms for frequencies above 3 kHz in (d)), the two standard amplification strategies resulted in different enhancements (panels (e),(f)) that only partially matched the reference NH response (b).}
\label{fig:AN_slope25}
\end{figure*}

\begin{table*}[htb!]
  \begin{center}
  \caption[Speech-in-noise performance across the different objective metrics (cf. Fig.~\ref{fig:optim}).]{Speech-in-noise performance across the different objective metrics (cf. Fig.~\ref{fig:optim}). Additional contrains were added to the best-performing loss functions to improve the performance of the DNN-HA models in noise. The considered objective metrics were computed for the 70-dB-SPL Flemish Matrix sentences at 2 SNRs (-12 and -6 dB). For each metric, bold font indicates cases for which the respective score was improved after processing.}
  \vspace{-10pt}
  \aboverulesep=0ex
  \belowrulesep=0ex
\resizebox{\textwidth}{!} {
  \begin{tabular}{l | c c c c c c c c c c c c}
\toprule
 & PESQ & STOI & ESTOI & SIIB (b/s) & SRMR & SNR\textsubscript{env} (dB) & sEPSM\textsuperscript{corr} & sCASP & HASPI & HASPI-HI & FADE (\%) & NRMSE (\%) \\
\toprule
\textbf{-12 dB SNR} & 1.0261 & 0.4633 & 0.1134 & 16.5122 & 0.4723 & 12.3064 & 2.5295 & 0.0953 & 0.0186 & 0.0209 & 14.83 & 26.3090 \\
\hline
$\La_{r}$ & 1.0233 & 0.4487 & 0.0981 & 10.9861 & \textbf{0.4822} & \textbf{13.1425} & 2.5197 & \textbf{0.0956} & 0.0170 & 0.0204 & \textbf{15.28} & \textbf{25.9403} \\
$\La_{r}$ + freq. emphasis & 1.0217 & 0.4549 & 0.1088 & 11.5469 & \textbf{0.4970} & \textbf{20.0793} & \textbf{2.5298} & \textbf{0.1031} & 0.0171 & 0.0183 & \textbf{16.06} & \textbf{25.9983} \\
$\La_{r}$ + $T_{r}$ & 1.0201 & 0.4491 & 0.0982 & 10.6441 & \textbf{0.4967} & \textbf{15.5072} & 2.5124 & 0.0934 & 0.0186 & \textbf{0.0234} & \textbf{15.33} & \textbf{26.0307} \\
\hline
$\La_{r,R}$ & \textbf{1.0262} & 0.4538 & 0.1035 & 12.0170 & \textbf{0.4879} & \textbf{12.5417} & 2.5267 & \textbf{0.0958} & 0.0186 & \textbf{0.0219} & \textbf{15.56} & \textbf{26.1802} \\
$\La_{r,R}$, complex STFT & 1.0240 & 0.4570 & 0.0987 & 12.2329 & 0.4646 & \textbf{12.9684} & 2.5203 & 0.0931 & 0.0171 & 0.0204 & \textbf{16.06} & \textbf{26.1978} \\
\hline
$\La_{r^2}$ & \textbf{1.0476} & 0.4409 & 0.0829 & 8.4139 & \textbf{0.4837} & \textbf{14.9675} & \textbf{2.5319} & \textbf{0.1005} & 0.0171 & 0.0203 & \textbf{16.17} & 26.5916 \\
$\La_{r^2}$ + $2000 \cdot \La_{estoi}$ & \textbf{1.0517} & 0.4391 & 0.0817 & 8.0005 & \textbf{0.4781} & \textbf{13.3881} & 2.5289 & \textbf{0.0970} & 0.0167 & 0.0201 & \textbf{15.50} & 26.5287 \\
$\La_{r^2}$, 12 layers & \textbf{1.0277} & 0.4349 & 0.0794 & 7.6450 & \textbf{0.4893} & \textbf{12.4560} & 2.5267 & \textbf{0.0956} & 0.0173 & \textbf{0.0211} & \textbf{15.78} & 26.6412 \\
$\La_{r^2}$, 10 layers & 1.0228 & 0.4316 & 0.0771 & 7.4117 & 0.4684 & \textbf{12.3924} & 2.5281 & \textbf{0.0955} & 0.0178 & \textbf{0.0210} & \textbf{16.33} & 26.4238 \\
$\La_{r^2}$, 41 CFs & 1.0249 & 0.4369 & 0.0787 & 8.3266 & 0.4570 & \textbf{12.4058} & 2.5263 & 0.0945 & 0.0176 & 0.0201 & \textbf{16.39} & 26.5635 \\
$\La_{r^2}$, noisy dataset & \textbf{1.0278} & 0.4439 & 0.0822 & 9.2533 & 0.4335 & \textbf{12.5311} & 2.5295 & \textbf{0.0957} & 0.0184 & \textbf{0.0217} & \textbf{16.33} & 26.6185 \\
$\La_{r^2}$, $T_{x}$ & \textbf{1.0433} & 0.4352 & 0.0794 & 7.6014 & \textbf{0.4794} & \textbf{12.8976} & 2.5221 & \textbf{0.0963} & 0.0177 & \textbf{0.0213} & \textbf{16.11} & 26.6320 \\
\hline
$\La_{r^2,R^2}$ & \textbf{1.0475} & 0.4265 & 0.0824 & 6.1092 & \textbf{0.5595} & \textbf{12.5572} & \textbf{2.5314} & \textbf{0.0961} & 0.0184 & \textbf{0.0210} & 14.78 & 26.4777 \\
$\La_{r^2,R^2}$, $T_{x}$ & \textbf{1.0355} & 0.4386 & 0.0886 & 8.4366 & \textbf{0.5114} & \textbf{14.4172} & 2.5184 & \textbf{0.0959} & 0.0156 & 0.0183 & 14.67 & \textbf{26.1061} \\
\hline
$\La_{r^2,r^2_p,R^2_p}$ & \textbf{1.0909} & 0.4035 & 0.0683 & 3.9637 & \textbf{0.5242} & \textbf{29.3251} & 2.4848 & 0.0877 & 0.0154 & 0.0171 & 13.89 & \textbf{26.0647} \\
$\La_{r^2,r^2_p,R^2_p}$, $T_{x}$ & \textbf{1.0381} & 0.3682 & 0.0540 & 2.4161 & \textbf{0.8709} & \textbf{63.2936} & \textbf{2.5366} & \textbf{0.1044} & 0.0144 & 0.0157 & 13.39 & \textbf{25.8803} \\
\bottomrule
\toprule
\textbf{-6 dB SNR} & 1.0234 & 0.5868 & 0.2625 & 69.5760 & 0.6708 & 71.4402 & 2.7906 & 0.1387 & 0.0536 & 0.0771 & 91.61 & 24.1906 \\
\hline
$\La_{r}$ & 1.0190 & 0.5608 & 0.2399 & 47.6385 & \textbf{0.7497} & 69.3254 & 2.7738 & 0.1387 & 0.0504 & 0.0751 & 84.89 & \textbf{22.9274} \\
$\La_{r}$ + freq. emphasis & \textbf{1.0306} & 0.5640 & 0.2494 & 49.3365 & \textbf{0.7652} & 67.1301 & 2.7656 & \textbf{0.1433} & 0.0517 & 0.0616 & 85.39 & \textbf{22.9693} \\
$\La_{r}$ + $T_{r}$ & 1.0205 & 0.5565 & 0.2385 & 46.3161 & \textbf{0.7374} & 66.1720 & 2.7077 & 0.1301 & 0.0333 & 0.0492 & 82.89 & \textbf{23.2096} \\
\hline
$\La_{r,R}$ & \textbf{1.0253} & 0.5745 & 0.2479 & 49.2899 & \textbf{0.6894} & 66.6963 & 2.7783 & 0.1367 & 0.0467 & 0.0645 & 85.06 & \textbf{23.3059} \\
$\La_{r,R}$, complex STFT & 1.0205 & 0.5800 & 0.2454 & 51.9126 & 0.6604 & 67.1233 & 2.7314 & 0.1292 & 0.0399 & 0.0571 & 84.61 & \textbf{23.3583} \\
\hline
$\La_{r^2}$ & \textbf{1.0318} & 0.5715 & 0.2209 & 40.5970 & \textbf{0.7093} & \textbf{72.6646} & \textbf{2.8197} & \textbf{0.1487} & \textbf{0.0609} & \textbf{0.0780} & 86.33 & \textbf{24.0329} \\
$\La_{r^2}$ + $2000 \cdot \La_{estoi}$ & 1.0206 & 0.5655 & 0.2183 & 39.5851 & \textbf{0.6940} & \textbf{71.6915} & 2.7853 & \textbf{0.1429} & \textbf{0.0577} & \textbf{0.0806} & 83.61 & \textbf{23.9403} \\
$\La_{r^2}$, 12 layers & 1.0204 & 0.5634 & 0.2152 & 38.4384 & \textbf{0.7218} & 69.8453 & 2.7852 & 0.1380 & \textbf{0.0573} & \textbf{0.0780} & 83.17 & \textbf{24.1602} \\
$\La_{r^2}$, 10 layers & 1.0221 & 0.5603 & 0.2153 & 41.5000 & 0.6579 & 71.0100 & 2.7883 & \textbf{0.1389} & \textbf{0.0572} & \textbf{0.0806} & 83.89 & \textbf{23.8140} \\
$\La_{r^2}$, 41 CFs & \textbf{1.0250} & 0.5671 & 0.2222 & 41.4595 & \textbf{0.6765} & 69.9502 & 2.7658 & 0.1356 & 0.0474 & 0.0680 & 87.28 & \textbf{23.9500} \\
$\La_{r^2}$, noisy dataset & 1.0208 & 0.5642 & 0.2178 & 44.4480 & \textbf{0.6799} & \textbf{71.6046} & 2.7862 & 0.1386 & \textbf{0.0571} & \textbf{0.0805} & 85.89 & \textbf{23.8263} \\
$\La_{r^2}$, $T_{x}$ & 1.0200 & 0.5647 & 0.2199 & 38.8743 & \textbf{0.7070} & 64.6625 & 2.7775 & \textbf{0.1409} & 0.0458 & 0.0596 & 83.89 & \textbf{24.0917} \\
\hline
$\La_{r^2,R^2}$ & \textbf{1.0241} & 0.5528 & 0.2208 & 33.8950 & \textbf{0.8065} & \textbf{72.0349} & \textbf{2.7985} & \textbf{0.1406} & \textbf{0.0574} & \textbf{0.0804} & 80.33 & \textbf{23.6901} \\
$\La_{r^2,R^2}$, $T_{x}$ & 1.0206 & 0.5547 & 0.2303 & 40.5785 & \textbf{0.7485} & 70.4307 & 2.7534 & \textbf{0.1419} & 0.0227 & 0.0265 & 79.78 & \textbf{23.3252} \\
\hline
$\La_{r^2,r^2_p,R^2_p}$ & \textbf{1.0394} & 0.4671 & 0.1716 & 17.3311 & \textbf{0.7549} & \textbf{76.4396} & 2.6522 & 0.1174 & 0.0233 & 0.0263 & 58.61 & \textbf{22.9323} \\
$\La_{r^2,r^2_p,R^2_p}$, $T_{x}$ & \textbf{1.0360} & 0.4138 & 0.1475 & 10.4308 & \textbf{1.1253} & 40.3225 & 2.7414 & 0.1240 & 0.0260 & 0.0219 & 57.06 & \textbf{22.6878} \\

\bottomrule
  \end{tabular}
}
  \label{tab:optim_all}
	\end{center}
	\vspace{-15pt}
\end{table*}

\begin{table*}[t!]
  \begin{center}
  \caption[Objective evaluation of speech in noise for two milder HI cases (cf. Fig.~\ref{fig:simpl}).]{Objective evaluation of speech in noise for two milder HI cases (cf. Fig.~\ref{fig:simpl}). For the trained models of the two HI profiles (Slope25 and 13,0,0), the considered objective metrics were computed for the 70-dB-SPL Flemish Matrix sentences in noise at 2 SNRs (-12 and -6 dB). For each metric, bold font indicates cases for which the respective score was improved after processing.}
  \vspace{-10pt}
  \aboverulesep=0ex
  \belowrulesep=0ex
\resizebox{\textwidth}{!} {
  \begin{tabular}{l | c c c c c c c c c c c c}
\toprule
\textbf{Slope25} & PESQ & STOI & ESTOI & SIIB (b/s) & SRMR & SNR\textsubscript{env} (dB) & sEPSM\textsuperscript{corr} & sCASP & HASPI & FADE (\%) & NRMSE (\%) \\
\toprule
-12 dB SNR & 1.0233 & 0.4621 & 0.1131 & 16.6771 & 0.4753 & 12.3064 & 2.5295 & 0.0953 & 0.0202 & 14.56 & 4.9053\\
\hline
$\La_{r^2,r^2_p,R^2_p}$, $T_{x}$ & 1.0225 & 0.4562 & 0.1087 & 13.9730 & 0.4701 & \textbf{12.4454} & 2.5264 & \textbf{0.0961} & \textbf{0.0210} & 14.56 & \textbf{2.4544} \\
NAL-RP \cite{byrne1990hearing} & 1.0232 & \textbf{0.4635} & \textbf{0.1158} & \textbf{16.9661} & \textbf{0.4819} & \textbf{12.9510} & 2.5252 & 0.0920 & 0.0193 & 14.56 & \textbf{3.1364} \\
Plack2004 \cite{ewert2011model,plack2004inferred} & 1.0231 & \textbf{0.4631} & 0.1102 & 14.7530 & \textbf{0.4955} & 10.3992 & 2.5121 & 0.0912 & 0.0184 & 13.00 & \textbf{3.5248} \\
\midrule
-6 dB SNR & 1.0215 & 0.5860 & 0.2621 & 69.1414 & 0.6735 & 71.4402 & 2.7906 & 0.1387 & 0.0744 & 92.44 & 5.6515\\
\hline
$\La_{r^2,r^2_p,R^2_p}$, $T_{x}$ & \textbf{1.0315} & 0.5693 & 0.2495 & 57.3403 & \textbf{0.6878} & 70.3828 & 2.7819 & 0.1387 & 0.0742 & 90.28 & \textbf{2.3236}\\
NAL-RP \cite{byrne1990hearing} & 1.0206 & \textbf{0.5885} & \textbf{0.2628} & \textbf{69.2732} & \textbf{0.7340} & 69.9391 & 2.7575 & 0.1265 & 0.0515 & \textbf{92.50} & \textbf{4.2178} \\
Plack2004 \cite{ewert2011model,plack2004inferred} & 1.0211 & \textbf{0.5878} & \textbf{0.2638} & 64.5134 & \textbf{0.8475} & 51.8560 & 2.7320 & 0.1224 & 0.0396 & 90.94 & \textbf{4.4447} \\
\bottomrule
\toprule
\textbf{13,0,0} & PESQ & STOI & ESTOI & SIIB (b/s) & SRMR & SNR\textsubscript{env} (dB) & sEPSM\textsuperscript{corr} & sCASP & HASPI & FADE (\%) & NRMSE (\%) \\
\toprule
-12 dB SNR & 1.0342 & 0.4619 & 0.1125 & 15.9589 & 0.4753 & 12.3064 & 2.5295 & 0.0953 & 0.0186 & 17.33 & 9.5387 \\
\hline
$\La_{r^2,r^2_p,R^2_p}$, $T_{x}$ & \textbf{1.0439} & 0.4571 & 0.1033 & 12.4507 & \textbf{0.5448} & \textbf{12.4930} & \textbf{2.5298} & \textbf{0.0959} & 0.0181 & 15.89 & \textbf{9.5285} \\
$\La_{r^2,r^2_p,R^2_p}$, $T_{x}$, 12 layers & 1.0342 & 0.4444 & 0.0976 & 9.7810 & \textbf{0.7456} & \textbf{12.6088} & \textbf{2.5305} & \textbf{0.0961} & 0.0186 & 14.67 & 9.7903 \\
\midrule
-6 dB SNR & 1.0340 & 0.5867 & 0.2607 & 68.3444 & 0.6756 & 71.4402 & 2.7906 & 0.1387 & 0.0535 & 92.94 & 7.9451 \\
\hline
$\La_{r^2,r^2_p,R^2_p}$, $T_{x}$ & 1.0197 & \textbf{0.5912} & 0.2581 & 57.2843 & \textbf{0.8714} & 69.9732 & \textbf{2.8099} & \textbf{0.1419} & \textbf{0.0546} & 89.72 & \textbf{7.2188} \\
$\La_{r^2,r^2_p,R^2_p}$, $T_{x}$, 12 layers & 1.0187 & 0.5733 & 0.2451 & 44.4043 & \textbf{1.6378} & \textbf{71.4853} & \textbf{2.8106} & \textbf{0.1422} & \textbf{0.0556} & 82.50 & \textbf{7.3899} \\
\bottomrule
  \end{tabular}
}
  \label{tab:simpl_all}
	\end{center}
	\vspace{-15pt}
\end{table*}

\begin{figure*}[htb]
\begin{center}
\includegraphics[width=\textwidth,height=\textheight,keepaspectratio]{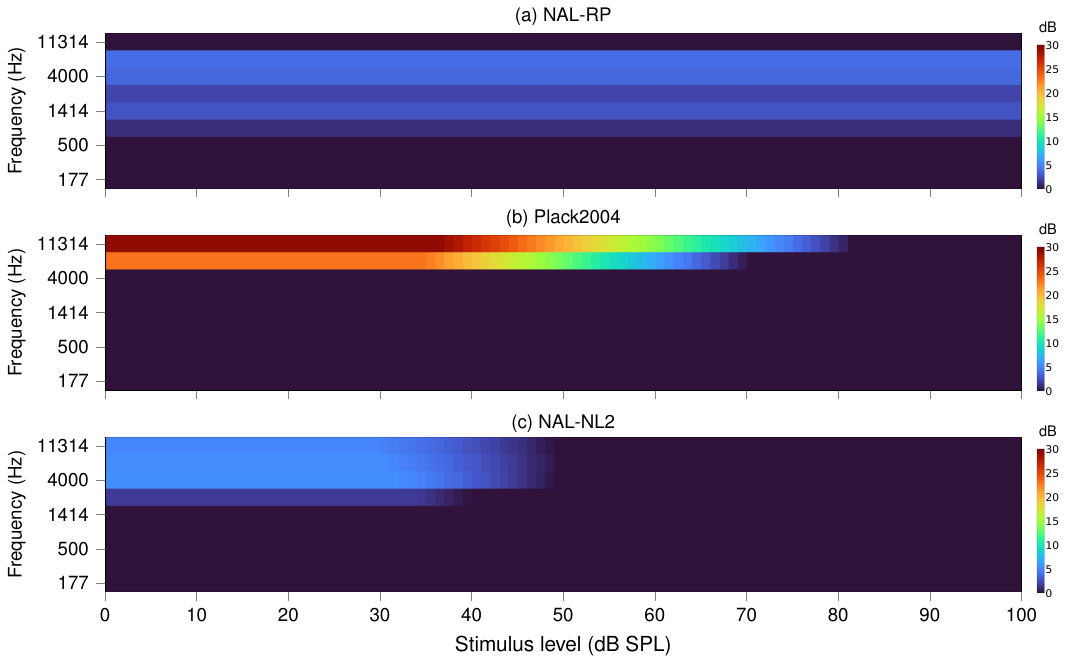}
\end{center}
\vspace{-10pt}
\caption[Gain tables for OHC loss (Slope25).]{Amplification provided by three state-of-the-art HA algorithms for OHC loss (Slope25). Each panel shows the gain (in dB) that was applied to an input stimulus across different frequency bands and instantaneous intensity levels by the openMHA toolbox \cite{kayser2022open}, based on the fittings of the three HA strategies. (a) The linear NAL-RP strategy \cite{byrne1990hearing} provided amplification of up to $\sim$4 dB to the input signal, with a constant gain applied across all stimulus levels. (b) The compressive Plack2004 strategy \cite{ewert2011model,plack2004inferred} provided amplification of up to $\sim$29 dB at low intensity levels and high frequencies, with the gain decreasing as the stimulus level increased to compress the dynamic range of the input. (c) The compressive NAL-NL2 strategy \cite{keidser2011nal} only amplified the input signal at low intensity levels and high frequencies, with gains of $\sim$5 dB for levels up to 40 dB SPL and frequencies above 4 kHz.}
\label{fig:gaintables}
\end{figure*}

\end{document}